\documentclass[fleqn]{mnras}

\PassOptionsToPackage{pdfpagelabels=false}{hyperref} 
\usepackage{hyperref} 
\usepackage{graphicx}	
\usepackage{amsmath}	
\usepackage{amssymb}	
\usepackage{color}
\usepackage{float}
\usepackage{ulem}
\usepackage{rotating}
\usepackage{booktabs}
\usepackage{tabularx}
\usepackage{longtable}
\usepackage{natbib}
\usepackage{xcolor}
\def\eq{\begin{equation}}
\def\en{\end{equation}}

\def\etal{{\it et al}\thinspace}

\def\apj{{\it Ap.J.}\thinspace}

\def\apjs{{\it Ap.J. Suppl.}\thinspace}
\def\aj{{\it A.J.}\thinspace}

\def\aap{{\it A\&A}\thinspace}

\def\mnras{{\it MNRAS}\thinspace}
\def\P3hat{{\mathaccent 94 P}_3}

\def\nat{{\it Nature}\thinspace}

\def\etal{{\it et al.}\thinspace}

\def\eg{{\it e.g.,}\thinspace}

\title [Pulsar B1237+25 Broadband A/R Emission Height Analysis]{Pulsar B1237+25 Aberration/Retardation Analysis from Decimeter to Decameter Wavelength: Challenge to ``Radius-to-Frequency Mapping''}

\author[Rankin, Zakharenko, Ulyanov, Kravtsov, Kumar, Grie{\ss}meier, Bhat, Wright, Weltevrede, \etal]
{Joanna M. Rankin$^{1}$, 
Vyacheslav Zakharenko$^{2}$, Oleg Ulyanov$^{2}$, Ihor Kravtsov$^{2}$, 
Pratik Kumar$^{3,4}$, 
\newauthor
Jean-Mathias Grie{\ss}meier$^{5,6}$, 
N. D. Ramesh Bhat$^{4}$, 
Geoff Wright$^7$, 
Patrick Weltevrede$^7$, 
\newauthor
Fabian Jankowski$^{5}$,
J\'er\^ome P\'etri$^8$
and
Gilles Theureau$^{5,6}$
\\
$^{1}$Physics Department, University of Vermont, Burlington, VT 05405, USA \\
$^2$Institute of Radio Astronomy of the National Academy of Sciences of Ukraine, Mystetstv St. 4, Kharkiv, Ukraine 61002 \\
$^3$Department of Physics and Astronomy, University of New Mexico, 210 Yale Boulevard NE, Albuquerque, NM 87106, USA \\
$^4$International Centre for Radio Astronomy Research, Curtin University, Bentley, WA 6102, Australia \\
$^5$LPC2E, OSUC, Univ Orleans, CNRS, CNES, Observatoire de Paris, F-45071 Orleans, France\\
$^6$ORN, Observatoire Radioastronomique de Nançay, Observatoire de Paris, Universit\'{e} PSL, Univ Orl\'{e}ans, CNRS, 18330 Nan\c{c}ay, France\\
$^7$Jodrell Bank Centre for Astrophysics, Department of Physics and Astronomy, University of Manchester, Manchester M13 9PL, UK \\
$^8$Universit\'e de Strasbourg, CNRS, Observatoire astronomique de Strasbourg, UMR 7550, F-67000 Strasbourg, France.}

\date{Accepted 24 September 2025. Received 06 May 2025}

\pubyear{2025}

\begin{document}
\label{firstpage}
\maketitle

\begin{abstract}
PSR B1237+25 is perhaps the canonical example of a pulsar with a core/double cone profile.  Moreover, it is bright with little spectral turnover, and its profile perhaps uniquely remains undistorted by scattering far into the decametric band.  Here we assemble more than a dozen of the highest quality profiles (30 MHz to 5 GHz) from half a dozen observatories, where possible polarimetric.  The pulsar's 2.6\degr\ core component marks the magnetic axis longitude, and we confirm that this point coincides both with the linear polarization angle inflection point and the zero-crossing of its antisymmetric circular signature---thus providing the possibility to estimate emission heights over a very broad band using aberration/retardation (A/R).  
We then carefully fit the profile components with Gaussians to identify and study the subtle asymmetries produced by A/R.  We find a consistent A/R in the pulsar's profiles of some 0.5\degr\ longitude or 2 ms---corresponding to a putative conal emission height of 200-400 km---with a formal error of about 100 km.  Our analysis finds no evidence whatsoever for an emission height increase with wavelength, the so-called ``radius-to-frequency mapping''.  Nor do we find any significant difference in A/R effect between the outer and inner cones.

\end{abstract}




\section{Introduction}
Pulsar B1237+25 has long provided the canonical example a core/double-cone (M) profile.  Its five components result from our sightline passing through both concentric emission cones as well as the central core beam over its 1.38-s rotation.  Its profiles show the characteristic narrow ``boxy'' form at high frequency that gradually open to a wide triple form at low frequencies \cite[\eg][hereafter Papers I and II, respectively ]{srostlik2005,smith2013}. The conal nature of its outer components is demonstrated by their nearly stationary in pulse phase (as opposed to drifting) 2.8-period subpulse modulation, and the core shows the usual antisymmetric circular polarization and characteristic width reflecting that of its polar cap (Paper II).  

\begin{figure}
\begin{center}
\includegraphics[width=70mm,angle=0.]{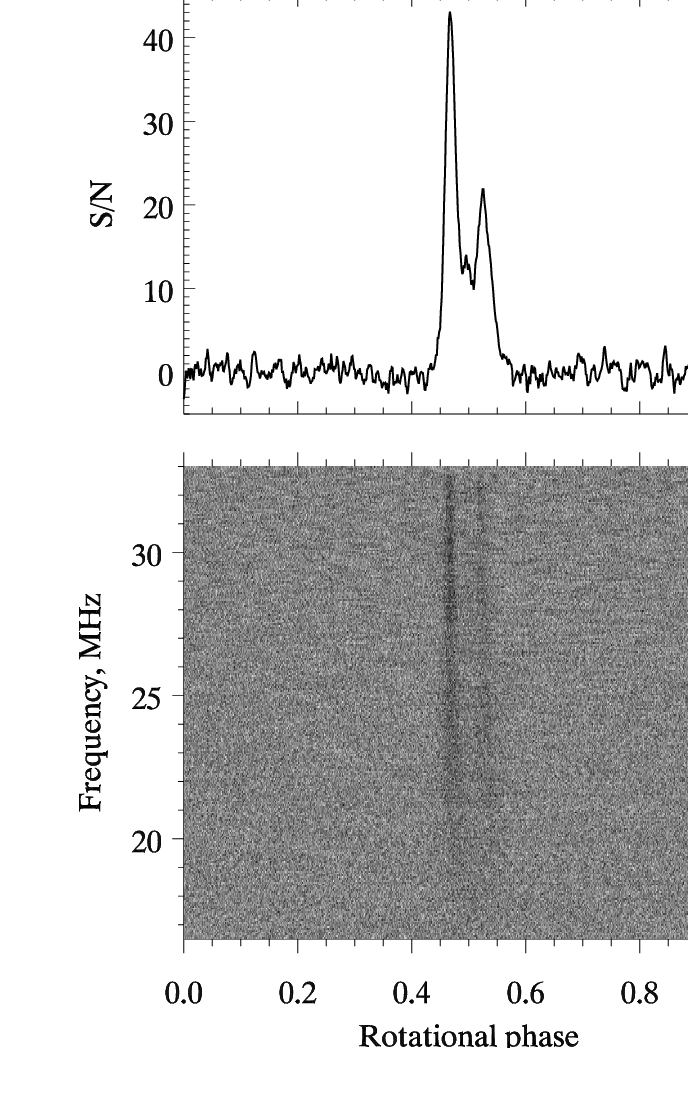}
\vspace{0.25in}
\caption{Broad band 16.5-33~MHz, full period B1237+25 profile from the Ukrainian UTR-2 telescope on 2021 November 24.}
\label{fig1}
\end{center}
\end{figure}

Moreover, B1237+25 is surely the brightest M pulsar to be observable at decametric wavelengths with virtually no scattering distortion\footnote{Scattering is well less than 1\degr\ longitude down to 30 MHz.} of its profile \citep{Zakharenko2013}. The UTR-2 telescope of the Institute of Radio Astronomy in Kharkiv, Ukraine\footnote {\url{https://old.nas.gov.ua/EN//Org/Pages/default.aspx?OrgID=0000544},\\ \url{https://en.wikipedia.org/wiki/Ukrainian_T-shaped_Radio_telescope,_second_modification}} has pioneered observations of the pulsar down to frequencies as low as 11 MHz. Fig.~\ref{fig1} shows a recent broadband profile of B1237+25 centred on about 25~MHz. As such, this pulsar provides a nearly unique opportunity to study the evolving structure of its profile over virtually the entire pulsar radio band below $\sim$5~GHz.  In particular, the aberration/retardation method \citep{BCW1991} can be used to estimate variations in emission height with wavelength so as to compare them with those estimated from the profile geometry.

The core/double-cone model was first suggested qualitatively by \cite{backer} and has since been developed quantitatively \citep[hereafter ETVIa,b]{ETVIb,ETVIa} and phenomenologically in a series of papers entitled {\it Toward An Empirical Theory of Pulsar Emission} \citep[\eg][hereafter ETXII]{ETXII}.  Not only do we now understand a good deal about the angular geometry of pulsar emission beams, we also have learned a good deal about the particular polarization and modulation characteristics of core and conal beams. Core emission appears dominant in pulsars with spindown energy loss rates $\dot E$ greater than about $10^{32.5}$ ergs \: $\text{s}^{-1}$ and conal below this \citep{Rankin2022}.  Within the polar fluxtube core energy flows centrally along the magnetic axis by contrast with the peripheral conal energy, and this seems to reflect two different (but often temporally intermixed) high and low energy source configurations of pair-plasma production per the particle-in-cell theory of \cite{Timokhin}.  The core/double-cone beam geometry has now been used to model large groups of pulsars in a recent series of papers in the {\it MNRAS} \cite[\eg][]{olszanski2022,RMpaperI} with very substantial success.

\S 2 considers the analysis of a multifrequency data set of Arecibo profiles from Hankins \& Rankin (\citeyear{hankins2010}) while introducing our aberration/retardation analysis, and \S 3  applies the analysis to a set of the best multiband profiles currently available.  \S 4 and 5 then provide a discussion and summary.

\section{Analysis of the Hankins \& Rankin Profiles}
\label{sec:HanRan}
The six-frequency set of B1237+25 time-aligned profiles using Arecibo observations shown in Figure~\ref{fig2} from the early 1990s was unusual for its time.  Most telescopes (including Arecibo) could not usually observe two simultaneous bands, so such a direct comparison of multi-band profiles could only be accomplished by aligning the profiles {\it as if} they had been observed simultaneously using accurate pulsar timing methods.  Moreover, most telescopes did not have Arecibo's frequency versatility over seven octaves.  

This said, the C-band observation at 4870 MHz was made prior to the installation of the Gregorian system so has a poorer signal-to noise ratio (S/N) due to using only a part of the instrument's collecting area.  Also, dedispersing methods were then relatively crude limiting the usable bandwidth at the lowest frequencies.

\subsection{Estimation of the Emission Height Using the Last Open Field Lines}

Backer's suggestion that pulsar emission beams entailed two concentric cones and a central core beam \citep[see their Fig.~1]{backer} immediately prompted efforts to model these beams using spherical geometry (\eg ETVIb: Fig. 2).

Two key angles describing the geometry are the magnetic colatitude (angle between the rotation and magnetic axes) $\alpha$ and the sightline-circle radius (the angle between the rotation axis and the observer’s sightline) $\zeta$, where the sightline impact angle $\beta$ = $\zeta-\alpha$. 
The three beams are found to have specific intrinsic angular dimensions at 1 GHz as a function of pulsar period: the core angular diameter being remarkably close to that of the polar cap {$\Delta_{PC}$} = $2.45\degr P^{-1/2}$ \citep[hereafter ETIV]{ETIV}, while the outside half-power radii of the inner and outer cones, {$\rho_{i}$} and {$\rho_{o}$} are $4.33\degr P^{-1/2}$ and $5.75\degr P^{-1/2}$ (ETVI).
Other studies such as \citet{gil}, \citet{KWJ}, \citet{Bhattacharya}, and \citet{mitra1999} have come to very similar conclusions. 

In practice, the magnetic colatitude $\alpha$ can be estimated from the width of the core component when present, as its half-power width at 1 GHz, $W_{\rm core}$ has empirically been shown to scale as {$\Delta_{\rm PC}/\sin\alpha$} (ETIV).  The sightline impact angle $\beta$ can then in turn be estimated from the steepest gradient of the linear polarization position angle (PPA) traverse $\chi_{0}$ (at the inflection point in longitude $\varphi$). $R$=$|d\chi/d\varphi|$ measures the ratio $\sin\alpha/\sin\beta$ (ETVIb, Eq.~3). Conal beam radii can similarly be estimated from the outside half-power width of a conal component or conal component pair at 1 GHz $W_{\rm cone}$ together with $\alpha$ and $\beta$ using Eq.~4 of ETVIb.  The characteristic height of the emission can then be computed assuming magnetic field dipolarity in the emission region\footnote{We also implicitly assume that the radial depth of the emission regions is small, such that the component widths reflect those of the emission beams.} using their Eq.~6. These 1-GHz inner and outer conal emission heights have typically been seen to concentrate around 130 and 220 km, respectively.

{\it However, this analysis depends on locating the conal emission adjacent to} the ``last open'' field lines (LoFL) at the polar fluxtube edges.  This LoFL approach has been widely used owing to the lack of another justifiable criterion, and it has seemed strange to treat the inner cone as well as the outer in this manner.  In any case, emission heights so estimated are geometric {\it characteristic} values and have no clear physical basis.  This phenomenological method of analysis has elevated the idea that radio emission heights are wavelength dependent to a broadly held dogma, known as ``radius-to-frequency mapping'' in pulsar jargon.  More physical emission heights can perhaps be estimated using aberration/retardation methods (see \citet{BCW1991} as corrected by \citet{dyks}) as we will explore below.

\begin{figure}
\begin{center}
\includegraphics[width=80mm,angle=0.]{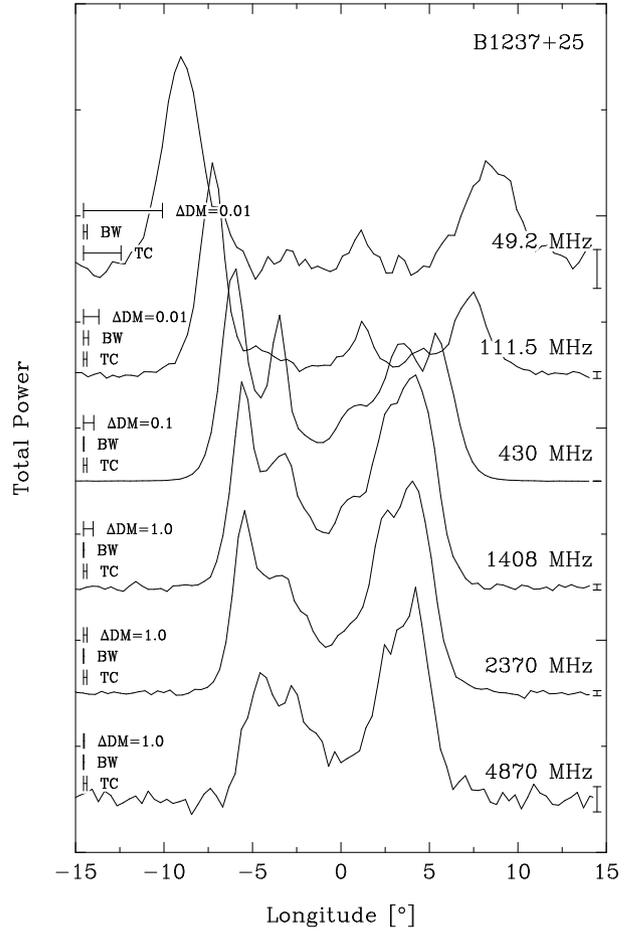}
\caption{Time-aligned Arecibo profiles of pulsar B1237+25 from \cite[][hereafter HR10]{hankins2010}. Here we see the evolution from the high frequency ``boxy'' form in which the weak core is not visible to the triple outer-conal form at low frequencies.  Note that the 49.2-MHz profile is broadened by a poorer time resolution.}
\label{fig2}
\end{center}
\end{figure}

Let us then examine what can be learned from the profiles in Fig.~\ref{fig2}.  Most obviously, a dramatic wavelength dependence can be seen in the outer conal component pair; their overall width nearly doubles in the seven octaves, and Fig.~\ref{fig1} exhibits the added increase of a further octave.  This effect was early identified by Jim Cordes (\citeyear{Cordes1978}) as ``radius-to-frequency mapping''\footnote{Cordes does allow for an alternative interpretation: ``An ad hoc counterproposal, however, is that all frequencies are radiated from the same narrow radial range by a broad-band mechanism but that the radiated spectrum varies systematically with distance from the magnetic pole. To reproduce the average-pulse-profile observations, the frequency of the spectral peak would necessarily decrease with distance from the magnetic pole. Neither theory nor observation can presently reject this counterproposal."}.  However, at high frequencies the widths seem to approach a minimum value. This behavior led Thorsett (\citeyear{Thorsett}) to suggest that the widths were best described by a constant plus a power law.\footnote{A detailed analysis of the pulsar's component widths and separations is given in \cite{ETVII}.} A recent study by \cite{Luo2025} apply these ideas to a large sample of pulsars using an inhomogeneous data set from the EPN database.

By contrast, the inner conal component pair positions change little down to the point where their relatively flat spectrum renders them indiscernible at 100 MHz and below.  Note also that the inner and outer conal component peaks do not merge at high frequency; they may become conflated but they do not overlap, suggesting that the minimum width of the outer cone is only a little larger than that of the inner cone at the highest frequencies---\citep[Fig.~A2]{olszanski2022}.  Finally, the core component is weakish in B1237+25, becoming conflated at high frequencies and at times appearing to have a different form at the lowest frequencies.  

By way of example, we model the LoFL emission heights using the HR10 profiles plus several others available in the literature in Table~\ref{OCLcFL}.  The frequency and the outside half-power widths $W$ are given in the first two columns, the outside conal radii $\rho$ and that computed from ETVIb: Eq.(4) $\rho_c$ in the third and fourth; $\beta/\rho$ in the fifth; and the estimated LoFL emission height $h_{LoFL}$ in the last column [Eq.(6)]. While the profile width doubles over this frequency range the estimated emission height more than triples.  The 206~km value at 1.4 GHz falls close to that found for many pulsars, this is a characteristic and phenomenological height estimate not a physical one.  

\begin{table}
\caption{B1237+25 Outer Conal Widths LoFL Emission Heights}
\label{OCLcFL}
\begin{center}
\begin{tabular}{cccccc}
\toprule
Freq &  $ W $ & $\rho$ &$\rho_c$ & $\beta/\rho$ & $h_{LoFL}$ \\
    (MHz) & (\degr) & (\degr) & (\degr) & & (km) \\
    \midrule
    \midrule										
4870 & 10.8 & 4.3 & 4.3 & -0.07 & 172 \\
2380 & 11.5 & 4.5 & 4.6 & -0.07 & 189 \\
1400 & 11.8 & 4.7 & 4.7 & -0.06 & 206 \\
 610 & 12.4 & 5.2 & 5.0 & -0.06 & 245 \\
 430 & 13.4 & 5.4 & 5.4 & -0.06 & 268 \\
 278 & 14.9 & 5.7 & 5.9 & -0.05 & 304 \\
 178 & 16.1 & 6.2 & 6.4 & -0.05 & 352 \\
 111.5 & 16.5 & 6.8 & 6.6 & -0.04 & 419 \\
 49.2  & 20.2 & 8.1 & 8.1 & -0.04 & 600 \\
    \bottomrule
\end{tabular}
\end{center}
The $\alpha$ and $\beta$ values here were taken as 53 and --0.53\degr, respectively.  
\end{table}

The Thorsett relationship was invented to describe the escalation in profile widths, but it can also be used to estimate the conal radius as a function of frequency.  We have used an equivalent relationship---that is, $\rho_c = \rho_0 [1 + (f/f_0)^\eta]$ to fit the computed $\rho$ values in the table and determine the parameters $f_0$, $\rho_0$ and $\eta$.  The values turn out to be some 69 MHz, 3.78\degr\ and --0.45. We also give $\rho_c$ values calculated using this Thorsett-like relation in Table~\ref{OCLcFL} for comparison.  

If the conal radiation really does come from sources immediately adjacent to the LoFLs (and we can assume dipolarity), then the $\rho_0$ value of 3.78\degr\ has significance and corresponds to a height of about 130 km---which in turn is a very typical height estimated for the inner conal emission.  We will come back to a discussion of this circumstance below.

\begin{figure*}
\begin{center}
\includegraphics[width=57mm,angle=0.]{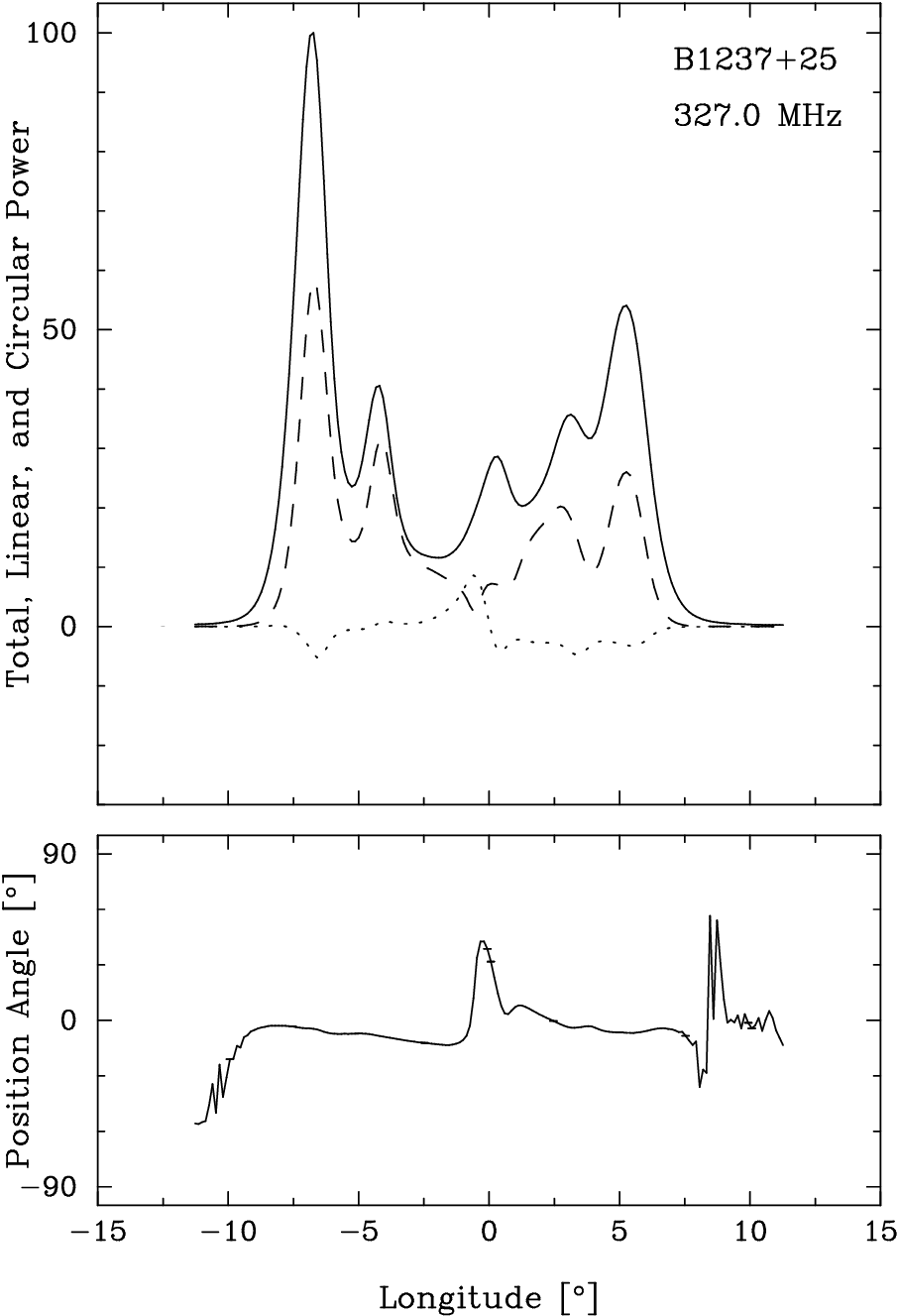}\ \ \ 
\includegraphics[width=57mm,angle=0.]{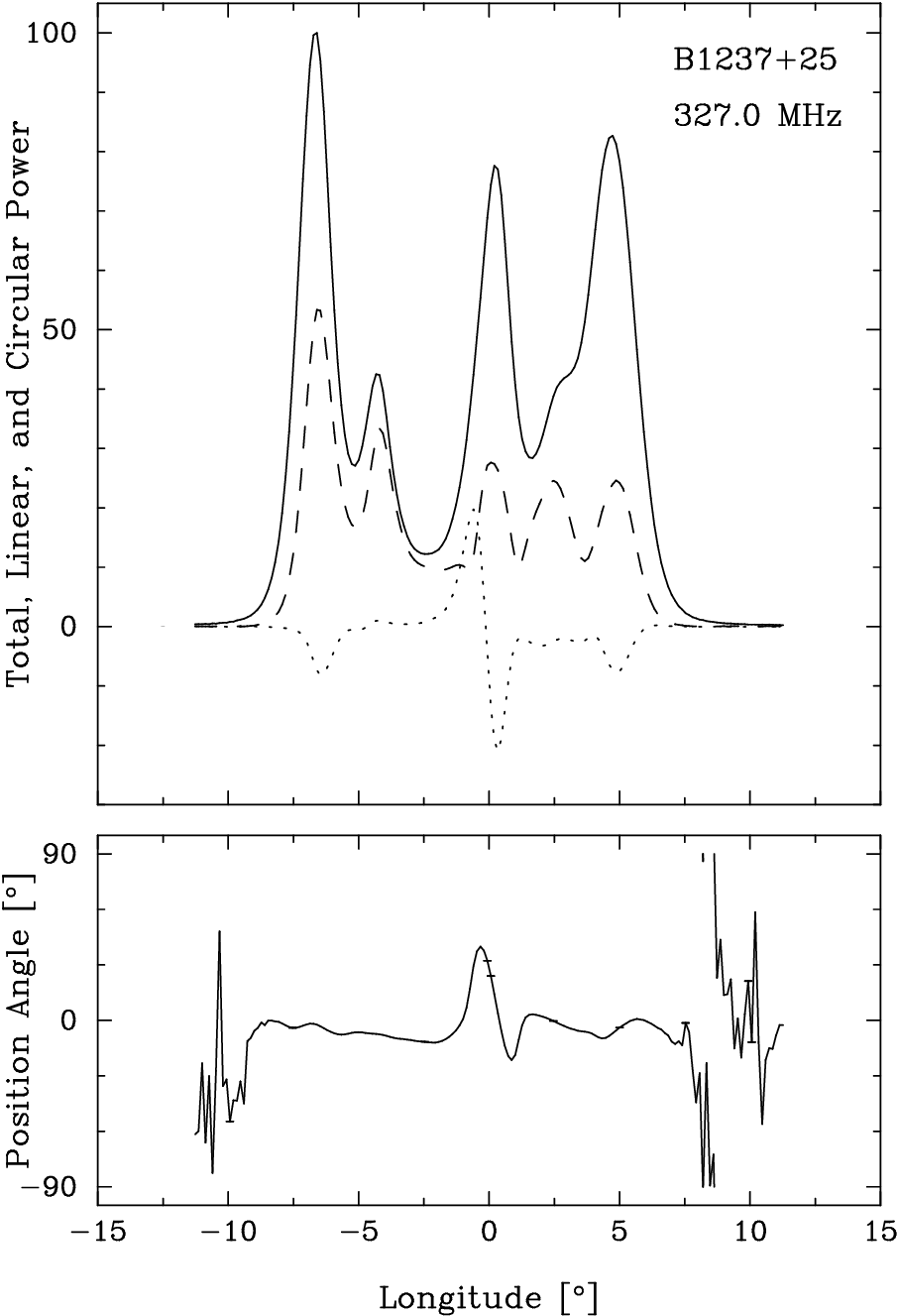}\ \ \ 
\includegraphics[width=57mm,angle=0.]{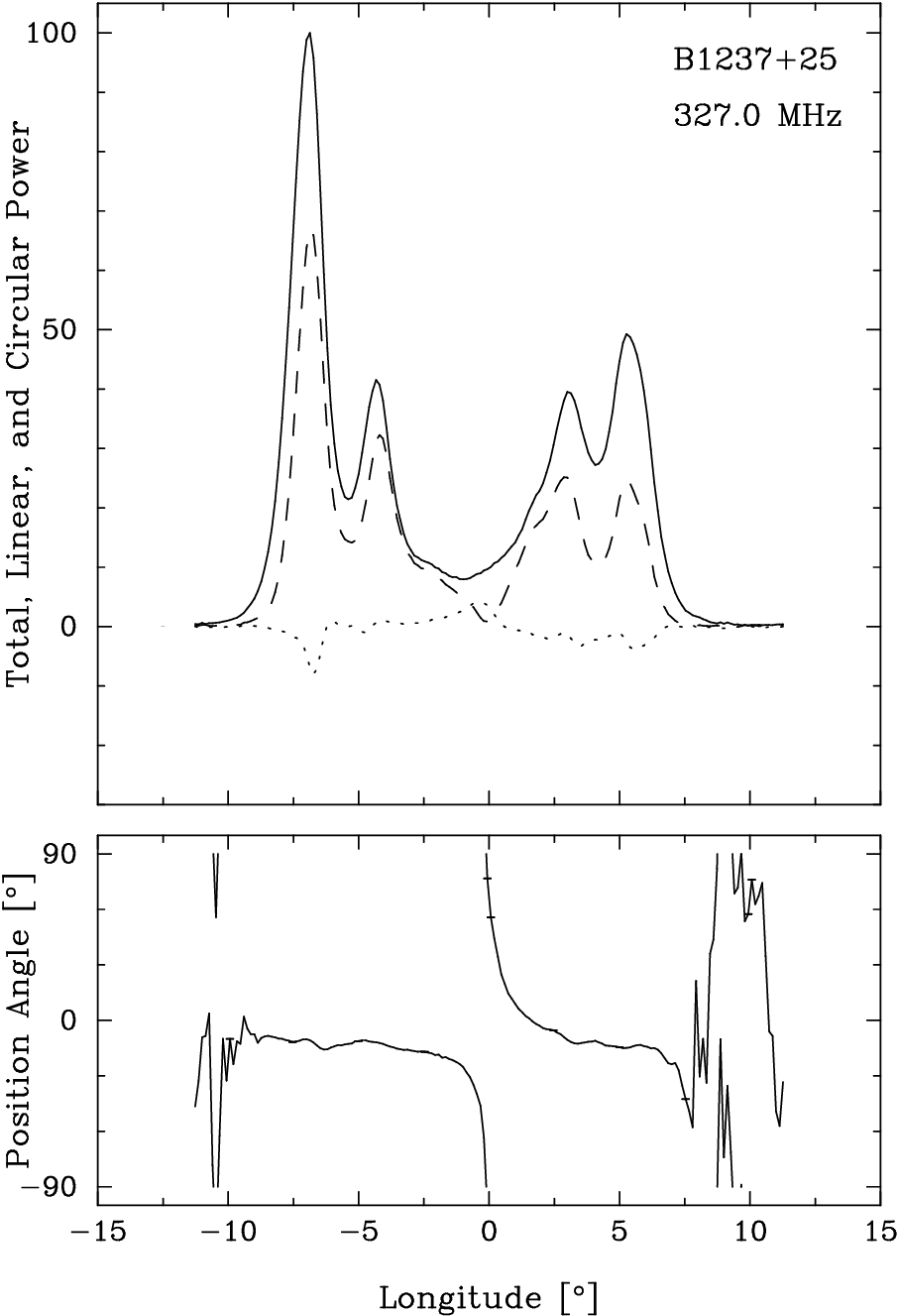}
\caption{Three B1237+25 profiles (Paper I): a) a typical 327-MHz total profile; b) flare-normal mode profile; and c) quiet-normal mode profile without core emission. Note that the core component center as marked by its antisymmetric $V$ in panel b coincides with the inflection point of the PPA traverse in panel c within 0.1\degr.  Here, the total power $I$, linear $L$ and circular $V$ polarization is plotted with solid, dashed and dotted curves in the upper panel and the PPA traverse in the lower panel. The 2-$\sigma$ off-pulse {\it RMS} bar at --10\degr\ is hardly visible because it is so small.}
\label{fig3}
\end{center}
\end{figure*}

\subsection{Estimation of the Emission Height Using Aberration/Retardation}
Blaskiewicz, Cordes \& Wasserman (\citeyear{BCW1991}, hereafter BCW) provided an important method for determining emission heights based on the aberration and retardation (A/R) that can be discerned in pulsar profiles.  Their analysis assesses the A/R as the time interval between the centerpoint of a conal component pair and the inflection point of a pulsar's linear polarization position angle (PPA) traverse.  Given that this fiducial longitude is often difficult to fix accurately and often coincides with the core component in profiles, it is convenient to use the core to label this point, although this assimilation has thus far not been fully established.  No obvious method is available to independently determine the height of the core emission.  The intrinsic core-component half-power width usually reflects the angular dimension of the polar cap (at the foot of the polar fluxtube), and this has suggested very low altitude emission---despite there being many arguments against this.  However, several well studied core components including that of B1237+25 (Paper II) exhibit intensity-dependent A/R suggesting emission heights of 100-200 km.  

Our primary purpose here is to show that the core center coincides with the PPA inflection point and can thus be taken as marking the fiducial longitude for purposes of using the above BCW method.  We explore this further in the next section.

Taking this approach, we can try to use the HR10 profiles to estimate the emission heights in a more physical manner. The profiles in Fig.~\ref{fig2} were time aligned with the centrepoint roughly marked by the core component $\varphi=0$. Then
\begin{equation}
h_{em} = c\frac{\Delta t}{2} = -c\frac{\varphi_c}{360\degr} \frac{P}{2}
        \label{eq1}
\end{equation}

Eq.~\ref{eq1} gives the emission height $h_{em}$ in terms of the time interval $\Delta t$ represented by the conal component midpoint in relation to its center as marked by the core component.\footnote{This form of the relationship reflects a correction by Dyks, Rudak \& Harding (\citeyear{Dyks2004}).}   

Turning now to the analysis, Table~\ref{OCA/R} gives the positions of the leading and trailing conal component half-power points $\varphi^i_l$ and $\varphi^i_t$ ($i=1,2$ for the inner and outer cones, respectively) relative to the putative core-component center, and their midpoint $\varphi^i_c$.  The time interval $\Delta t$ is given in the fifth column and the corresponding emission height 
$h^i_{em}$ in the rightmost column. 

\begin{table}
\setlength{\tabcolsep}{4pt}
\caption{A/R Height Analysis of the Hankins \& Rankin widths}
\label{OCA/R}
\begin{center}
\begin{tabular}{rccccc}
\toprule
Freq &  $\varphi^i_l$ & $\varphi^i_t$  & $\varphi^i_c $ & $\Delta t^i$ &  $h^i_{em}$  \\
Cone & (\degr) & (\degr) & (\degr) & (ms) & (km) \\
    \midrule
    \midrule										
4870/2 & -6.0$\pm$0.2 & 5.1$\pm$0.2 & -0.5$\pm$0.1 & -1.7$\pm$0.5 & 259$\pm$110 \\
2370/2 & -6.5$\pm$0.2 & 5.5$\pm$0.2 & -0.5$\pm$0.1 & -1.9$\pm$0.5 & 288$\pm$110 \\
1408/2 & -6.5$\pm$0.2 & 5.6$\pm$0.2 & -0.5$\pm$0.1 & -1.7$\pm$0.5 & 259$\pm$110 \\
430/2 & -7.5$\pm$0.2 & 6.5$\pm$0.2 & -0.5$\pm$0.1 & -1.9$\pm$0.5 & 288$\pm$110 \\
111.5/2 & -8.5$\pm$0.2 & 7.7$\pm$0.2 & -0.4$\pm$0.1 & -1.5$\pm$0.5 & 230$\pm$110 \\
49/2 & -10.5$\pm$0.2 & 9.7$\pm$0.2 & -0.4$\pm$0.1 & -1.5$\pm$0.5 & 230$\pm$110 \\
\\										
4870/1 & -4.2$\pm$0.3 & 3.7$\pm$0.3 & -0.3$\pm$0.2 & -1.0$\pm$0.8 & 144$\pm$165 \\
2370/1 & -4.1$\pm$0.3 & 3.5$\pm$0.3 & -0.3$\pm$0.2 & -1.2$\pm$0.8 & 173$\pm$165 \\
1408/1 & -4.1$\pm$0.3 & 3.6$\pm$0.3 & -0.3$\pm$0.2 & -1.0$\pm$0.8 & 144$\pm$165 \\
430/1 & -4.3$\pm$0.3 & 3.7$\pm$0.3 & -0.3$\pm$0.2 & -1.2$\pm$0.8 & 173$\pm$165 \\ 
111.5/1 & n/a									\\
49/1 & n/a									\\
\bottomrule
\end{tabular}
\end{center}
Notes: "2" and "1" label the outer and inner cones, respectively.  $\varphi^i_l$ and $\varphi^i_t$ are the leading and trailing conal component positions; 
$\varphi^i_c$ their centers; $\Delta t^i$ the total A/R shifts in ms; and $h^i_{em}$ the A/R emission heights computed from Eq.~\ref{eq1}.  The measurement errors are taken to be 0.2 and 0.3\degr\ for the outer and inner conal components, respectively. 
\end{table}

This is one of the very first instances that we have had the opportunity to apply the A/R technique in a broad multifrequency context.  It has been applied in a number of cases to profiles at frequencies around 1 GHz, and there the results have often given height values agreeing within a factor of 2 or so with the LoFL values.  Here we can see in Table~\ref{OCA/R} that the outer cone A/R height value around 1 GHz is roughly compatible with the LoFL value above.  However, their frequency evolution is strikingly contrary to what was seen from the LoFL values in Table~\ref{OCLcFL}.  The outer conal A/R heights do not seem to vary within their errors, and this seems to be so for the inner cone as well.

The analysis we are presenting here is a naive one.  A number of doubts can be voiced.  It is utterly remarkable that the results are as indicative as they appear to be.  B1237+25 is not a fast pulsar.  With an emission height of the order of 100 km, a retardation of 3 ms would be hardly different than 1\degr\ of B1237+25 longitude---only a few times larger than our measurement uncertainties. The S/N of some profiles is poor.  The fiducial longitude in relation to the core is poorly established.  Need we go on .... ? 

{\it This all said, were the physical emission height increasing with wavelength as indicated by the LoFL geometry, the naive A/R analysis in Table~\ref{OCA/R} would seemingly have been able to show it.}

Against the above, A/R is such a small effect in ordinary pulsars that some colleagues have opined that it could not be reliably estimated.  In millisecond pulsars the A/R effect is potentially even more substantial, and it appears to be a significant factor in the profile structure of several such objects with core/cone emission beams, \eg J0337+1715 (ETXII).  Three published studies have attempted to estimate A/R emission heights in B1237+25 (\cite{gg2003}, Paper I and \cite[see Appendix]{KrzeszowskiMitra2009} which we will assess and critique below.

\subsection{Does B1237+25's Core Component Define the Fiducial Longitude?}
As can be seen in Fig.~\ref{fig2}, B1237+25's central core component is weak and appears to have an inconsistent position and width \citep[\eg see][Fig.~A2]{olszanski2022} within the pulsar's profiles.  If so, any serious A/R analysis would be impossible, therefore a comprehensive study of its core is required.  

A further complication is that B1237+25's profiles usually represent an indeterminate average of its three emission modes.  Most average profiles include segments of its unusual abnormal mode, but all include both the normal and flare-normal intervals.  All three were studied in detail at 327 MHz in Paper I, and a key insight is that subpulses associated with the core are not usually continuous:  in the normal mode they occur in ``flares'' typically every 50-60 pulses, while abnormal intervals do show bright core emission in continuous sequences of a few to several hundred pulses.  

Figure~\ref{fig3} shows three 327-MHz profiles from Paper I.  The first on the left provides a typical total profile.  The pulse sequence here included normal mode as well as some abnormal.  Intervals of flare-normal mode were then averaged to provide the middle profile.  Note its strong, slightly asymmetric core component showing 25\% antisymmetric circular and a linear polarization mirroring the total power.  The mostly flat PPA traverses show an interruption under the core rather than the expected steep traverse for a pulsar with such a small $\beta$ value.  However, the profile on the right, consisting of mainly conal power, shows no core feature and one of the steepest PPA inflections known.  Careful inspection of the latter plots further shows that the Stokes $V$ zero crossing is coincident with the PPA inflection point within some 0.1\degr.  Note that the PPA inflection does not align with the core peak due to its (variable) asymmetry; the $V$ zero crossing point provides a more reliable marker.   

For a pulsar with a 1.38-s rotation rate and an $\alpha$ of 54\degr, we expect an intrinsic core of some 2.1\degr\ and a profile width of 2.6\degr---both reflecting the dipolar polar-cap angular width.  The core in the middle plot is a little narrower and asymmetric---something often but not always seen in B1237+25.  

We therefore conclude that B1237+25's core component is correctly identified and exhibits the expected properties, but its average profiles are distorted by the admixture of the three modes as well perhaps by intensity-dependent A/R (Paper II; Fig.~11).  The four PPA inflections under the core speak to the complexity of its polarization-modal composition as was studied in Papers I and II. The PPA profiles in Fig.~\ref{fig3} are absolute (measured counterclockwise from North on the sky).  The inflection near $\pm$90\degr\ (rightmost) together with the pulsar's proper motion direction identifies the profile as primarily X-mode; whereas in the average and flare-normal profiles the core is depolarized by slightly stronger O-mode radiation (\citealt{Johnston+2005}; \citealt[hereafter ETXI]{ETXI}).\footnote{Owing to this complexity, the values in \cite[Table I]{ETXI} incorrectly report \cite{Johnston+2005}'s values for PA$_0$ and $\Psi$ of --66 and 1\degr, respectively.}

\begin{figure*}
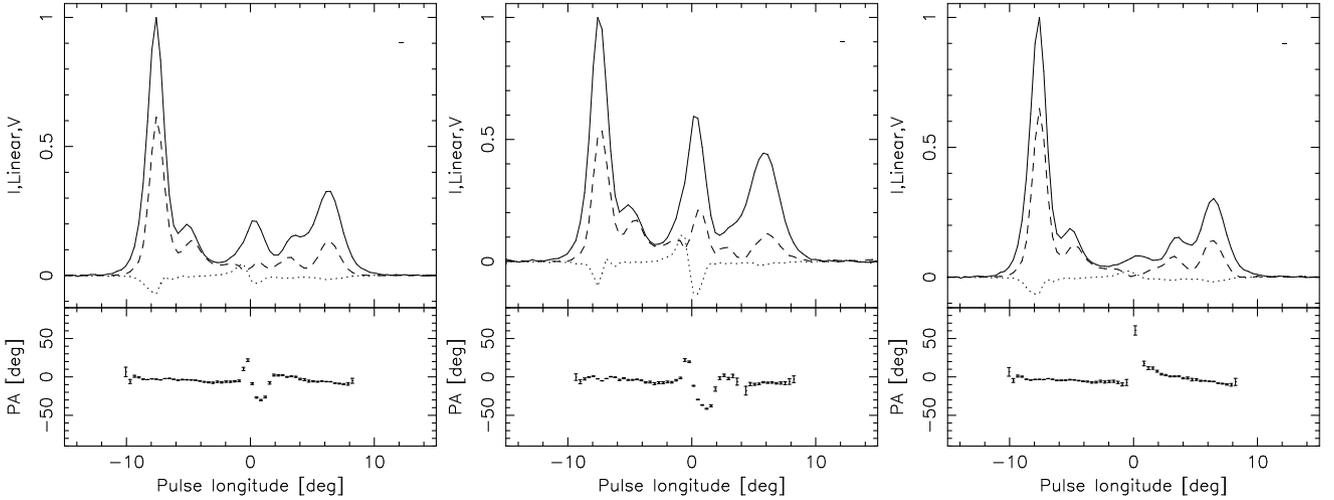

\begin{center}
\includegraphics[width=65mm,angle=270.]{plots/L78449_all.ps}
\includegraphics[width=65mm,angle=270.]{plots/L78449_bright_core.ps}
\includegraphics[width=65mm,angle=270.]{plots/L78449_weak_core.ps}
\caption{Polarimetric average profile of pulsar B1237+25 at 151 MHz from LOFAR \citep{Noutsos2015}. This figure can be compared to Fig.~\ref{fig3}, Which shows very similar features.  {\em Left panels:} The total profile with $I$, $L$ and $V$ shown as solid, dashed and dotted curves in the top panel. The resolution and off-pulse noise level (although so small that to be barely visible) is shown by the boxes in the top-right corner. The PPA traverse is shown in the bottom panel.  {\em Middle panels:}  Profile averaging 5-second long sub-integrations with strong core emission.  {\em Right panels:}  Profile with weak core emission. The PPA is only plotted when $L > 2\sigma$.}
\label{fig4}
\end{center}
\end{figure*}

\begin{figure*}
\begin{center}
\includegraphics[width=57mm,angle=-0.]{plots/pB1237+25.20230719_70h.ps}
\includegraphics[width=57mm,angle=-0.]{plots/pB1237+25.20230719_50h.ps}
\includegraphics[width=57mm,angle=-0.]{plots/pB1237+25.20230719_30h.ps} \\
\vspace{0.25in}
\includegraphics[width=72mm,angle=+90.]{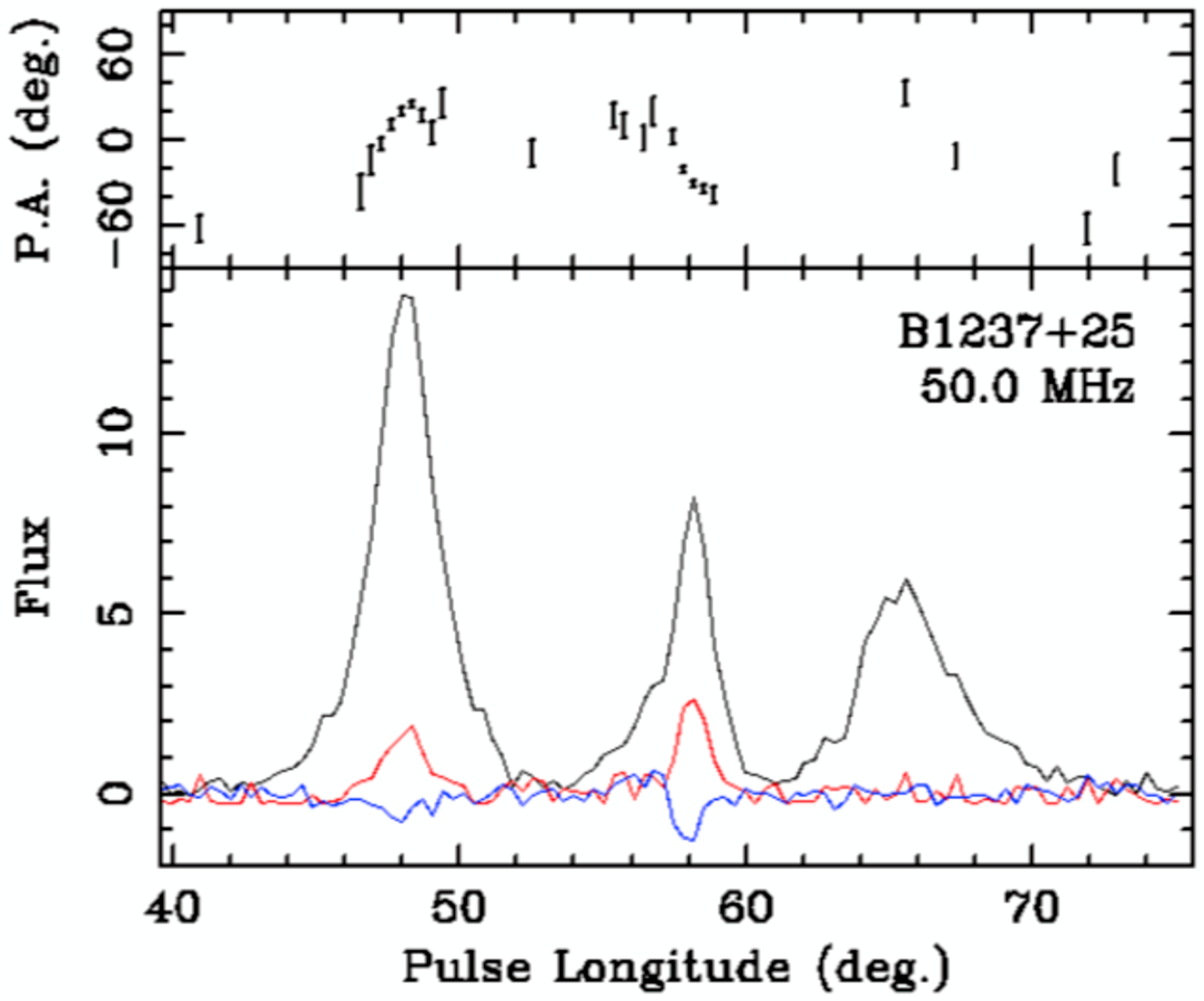}
\includegraphics[width=72mm,angle=+90.]{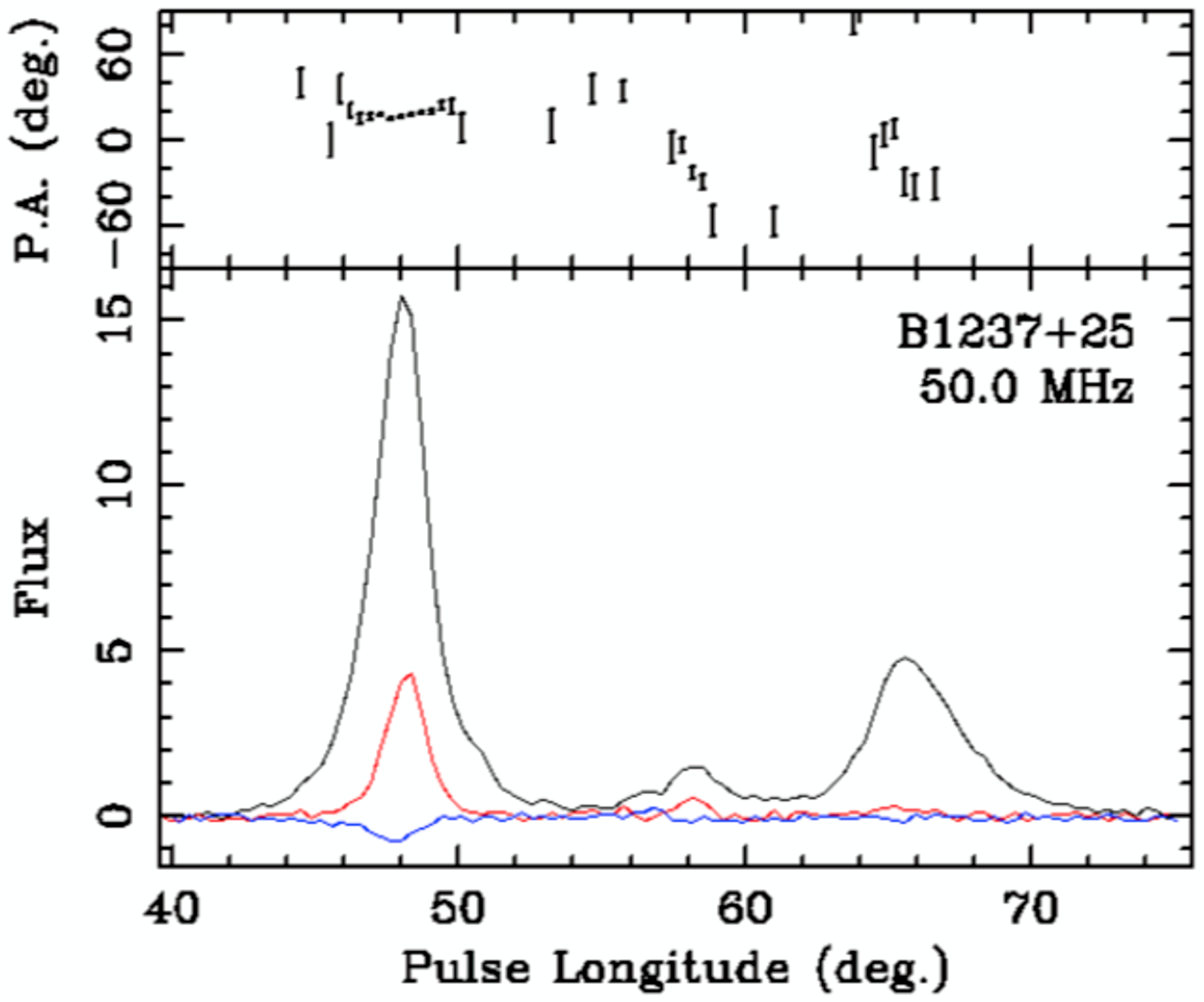}
\caption{{\em Upper panels:} Simultaneous polarimetric observations at 70-, 50-, and 30-MHz  with NenuFAR. Solid black line: total intensity $I$. Dashed green and dotted magenta lines: $L$ and $V$.  Note that each frequency shows a core component with a stronger trailing portion and that the core becomes ever more depolarized at low frequency. The resolution and off-pulse noise level are shown by the small boxes at --20\degr, and the PPA gaps occur due to regions where $L < 2\sigma$.
{\em Lower panels:} NenuFAR 50-MHz flare- and quiet normal profiles as in Figs. \ref{fig3} and \ref{fig4}.}
\label{fig5}
\end{center}
\end{figure*}

\begin{figure}
\begin{center}
\includegraphics[width=57mm,angle=0.]{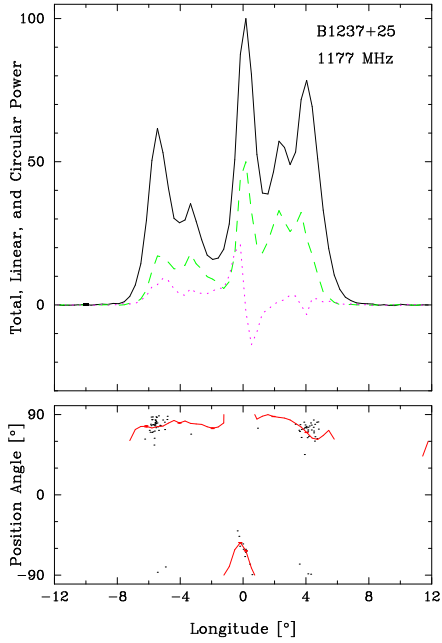}
\caption{Polarimetric flare-normal mode profile of B1237+25 at 1177 MHz \citep{Olszanski2019}.  The total power, $L$ and $V$ are shown in black, green and purple in the upper panel and the PPA traverse in the lower panel.  This Arecibo normal mode sequence of 672 pulses shows about 46 flare-normal pulses.  Note as in Fig.~\ref{fig3} (middle panel) that the core's asymmetry reflects a missing leading portion.}
\label{fig6}
\end{center}
\end{figure}

A further important question is whether the core-component structure in relation to the PPA traverse remains fixed as a function of frequency.  Polarimetry at lower frequencies becomes increasingly difficult, but Fig.~\ref{fig4} shows the same high quality 151-MHz data as in \cite{Noutsos2015}---and its PPA traverse appears nearly identical to that in Fig.~\ref{fig3}. Furthermore, utilizing that the data was recorded with 5 second sub-integrations (between 3 and 4 pulses per sub-integration), allows a meaningful segregation in emission where the core component is bright (26 per-cent of the duration of the observation) and weak (74 per-cent). Indeed, the panels in Fig.~\ref{fig4} show very similar behavior compared to Fig.~\ref{fig3}. Further, Fig.~\ref{fig5} shows the increasingly depolarized NenuFAR profiles into the decametric band that appear compatible as well as 50-MHz mode separations (from 10.7-s subaverages) similar to those in Figs.~\ref{fig3} and \ref{fig4}.  In particular, the 70-MHz one may show a part of the very steep PPA traverse. At the other end of the spectrum, the 1177-MHz flare-normal profile in Fig.~\ref{fig6} shows again a similar feature two octaves higher compared to Fig.~\ref{fig3}.

\section{Aberration/Retardation Analysis of High Quality B1237+25 Multiband Profiles}

The above discussion suggests that the radio emission heights in pulsar B1237+25 do not vary with frequency, and if this is so it contraverts the longstanding appearance that outer conal emission exhibits the wavelength dependence known as ``radius-to-frequency mapping''.  As we have noted, the above A/R analysis raises a number of concerns: some of the profiles are poorly measured, only widths were analyzed not component peaks, and an unvarying A/R could mask other issues.  Both issues depend strongly in B1237+25 on the structure of its core component and how it is treated and interpreted in the analysis.  In the following, we use high quality observations across a wide frequency band to improve upon the above analysis.

Our purpose in what follows is to use the physical effects of A/R (Blaskiewicz, Cordes \& Wassermann 1991) {\it to estimate} emission heights over the largest possible band.  This is the novel, primary and we believe significant part of our effort.  Given that the A/R method has a strong physical foundation---that its application here is novel---and that LoFL methods are standard and represent an unverified phenomenological interpretation, we will give strong priority to the former.  This all said, we also emphasize that the A/R analysis assumes the magnetic field dipolarity that we believe obtains at intermediate heights in the pulsar magnetosphere.

\subsection{Observations}
The low frequency radio observatories that have contributed observations to this project are given in Table~\ref{OBS}, together with the frequency range that was used and references to published work where possible.  

\begin{table}
\caption{Multiband Observations
}
\label{OBS}
\begin{center}
\begin{tabular}{ccc}
\toprule
    Observatory &Freq Range & References and notes  \\
    Code & (MHz) &  \\
    \midrule
    \midrule	
UTR-2 & 16.5-33 & war damaged$^\dagger$  \\ 
Ukraine & &         \\
NenuFAR & 20-80 & this work  \\
France \\
LWA     & 35-80 & \cite{Kumar2025}  \\
New Mexico \\
HBA     & 40-80 & \cite{Noutsos2015}  \\
Netherlands \\
LBA     & 120-180 & \cite{Bilous2016}  \\
Netherlands \\
MWA     & 140-170 & \cite{Bhat2023} \\
Australia \\
AO   & 327, 1177, 4400 & \cite{Olszanski2019} \\
Puerto Rico \\

\bottomrule
\end{tabular}
\end{center}
Notes: $^\dagger$UTR-2 was partially destroyed by Russian occupiers in 2022. Nevertheless, the NAS of Ukraine and the Institute of Radio Astronomy of the NAS of Ukraine are working to preserve the surviving parts of the UTR-2 radio telescope and have begun its reconstruction.  {\it Science} 14 July 2024, doi: 10.1126/science.zu5np4h \\
Observatory codes: AO, Arecibo Observatory; HBA, LOFAR High Band Antenna; LBA, LOFAR Low Band Antenna; LWA, Long Wavelength Array; MWA, Murchison Widefield Array; NenuFAR, NenuFAR Telescope; and UTR-2, Ukrainian T-shaped Radio telescope, second modification.
\end{table}

\begin{figure*}
\begin{center}
\includegraphics[width=150mm,angle=0.]{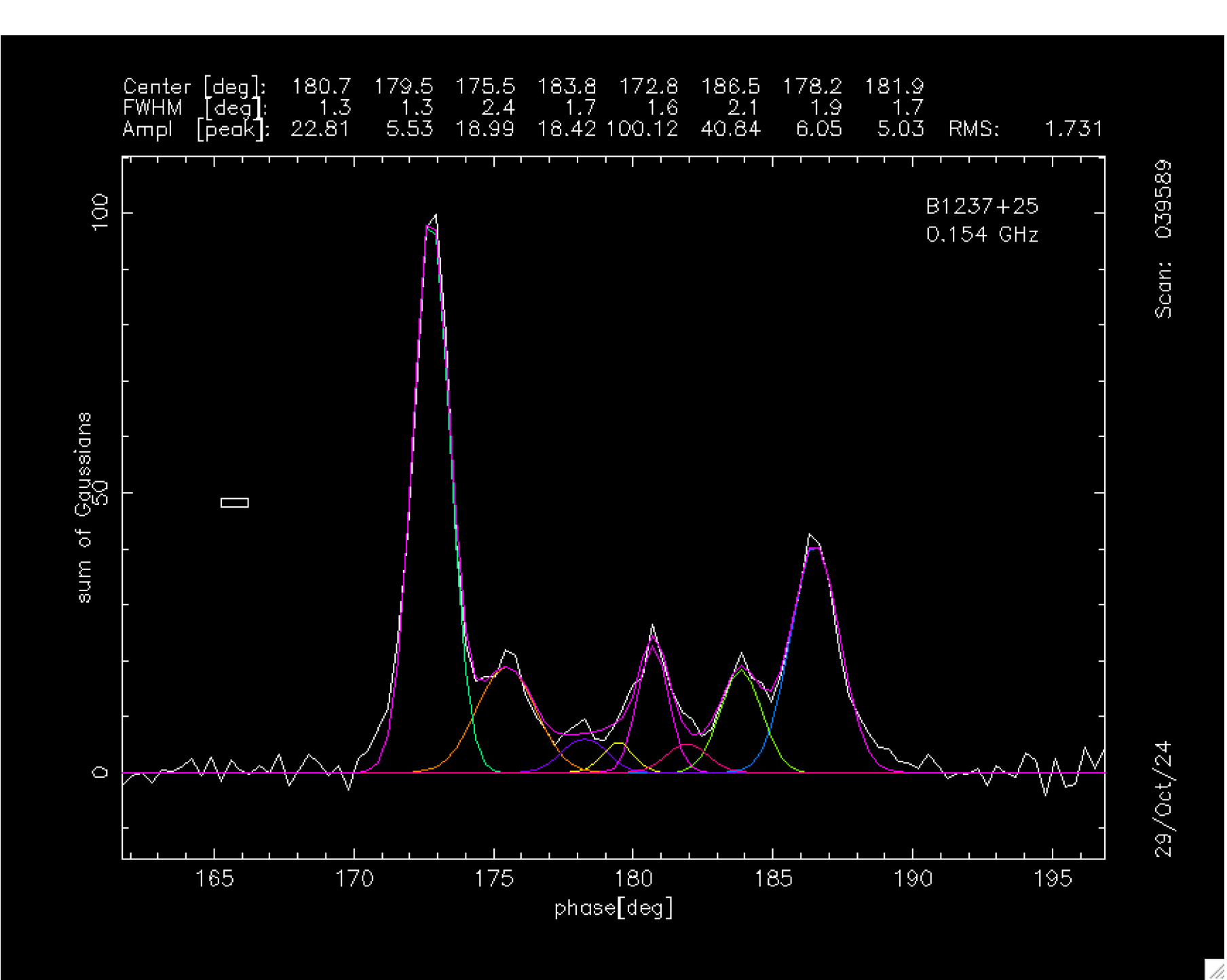}
\caption{Gaussian-fitted MWA profile of pulsar B1237+25 at 154.24 MHz.  The white curve shows the observation.  Each of the 8 Gaussians are shown in different colors with the dotted blue line giving their sum and the dash-dotted green showing the difference between the observation and the fitted function.  The core is represented by two Gaussians as discussed above; note that the leading one is plotted in yellow and a bit difficult to discern.  The residual diagram is given in the righthand panel of Fig.~\ref{figB6}}.   
\label{fig8}
\end{center}
\end{figure*}

\subsection{Modeling B1237+25's Core and Conal Emission}
First, we will model B1237+25's profile using Gaussian components.  This will permit computing A/R heights using both the peaks and widths of the component pairs.  Such modeling has long been practiced with considerable success following its introduction by Michael Kramer (\citeyear{kramer94})\footnote{Gaussian fitting was introduced in a slightly different pulsar context by \cite{foster1991}}.  Experience has shown that pulsar emission components often show a near-Gaussian form \citep{kramer+94}, usually departing mainly around their 10\%-level ``wings''.  

Second, the latter two modal profiles in Fig.~\ref{fig3} show that the PPA inflection point is accurately aligned with the center of the core component as well as with the Stokes $V$ zero-crossing point at 327 MHz.  No similarly detailed analysis is available at other frequencies, but the similarities of the total profiles in other bands (Fig.~\ref{fig4}; 151 MHz, and Fig.~\ref{fig5}; 30, 50 and 70 MHz) strongly suggests that this is so overall.  Fig.~\ref{fig4} attempts a normal-mode segregation similar to that in Fig.~\ref{fig3}, but is limited by the observation's 5-s averaging.  This gives strong evidence that the fiducial longitude aligns with the core component center.

\begin{figure}
\begin{center}
\includegraphics[width=57mm,angle=0.]{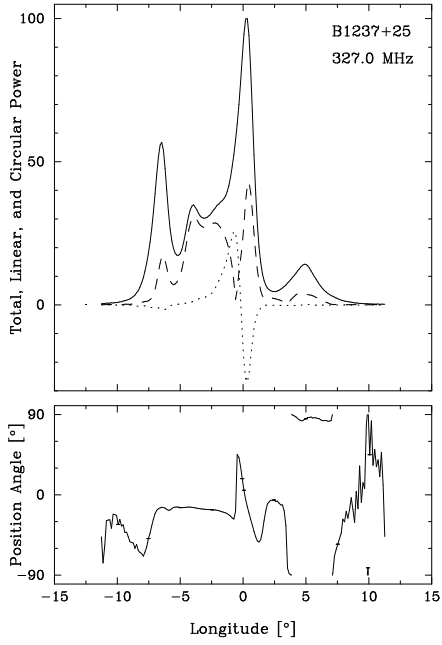}
\caption{B1237+25 abnormal mode profile (Paper I).  Note the greatly larger power level between the leading inner conal component and the core as well as the single trailing component instead of a pair.  Also interesting is the bright core's balanced antisymmetric $V$ (dotted line) but asymmetric $L$ (dashed line) due to a depolarized leading portion.}
\label{fig7}
\end{center}
\end{figure}

Third, the variability and asymmetry of form exhibited by B1237+25's core component requires special attention.  Its core component is exceptional as most have a nearly Gaussian form. In most observations as we have seen above (\eg Figs. \ref{fig3}, \ref{fig4}--\ref{fig6} and \ref{fig8}), the core component is found to be non-Gaussian with a shallow leading and steep trailing edge, such that leading power needed to give it a near Gaussian form is missing.  Its half-power width is then somewhat less than the full 2.6\degr\ that is at times observed.  As seen in the flare-normal profile of Fig.~\ref{fig3}, the form of the core sometimes approaches a symmetrical Gaussian form but only rarely assumes it.  Why this is so is not fully clear, but Paper I makes a strong case that it can be traced to the admixture of modes in B1237+25 total profiles.  

Note also that B1237+25 total profiles usually exhibit a double inflection of the PPA traverse coinciding with the core component (\eg central panel of Fig.~\ref{fig3}).  This unusual behavior was also studied in Paper I and traced to a dominant secondary (O) polarization mode associated with the core emission.  When this power is absent in a totally conal profile, the full PPA traverse is exhibited as in the rightmost panel of Fig.~\ref{fig3}.

Given that the trailing portion of B1237+25's core is seemingly always present, the whole can be well modeled by two adjacent Gaussians of 1.3\degr\ half-power width spaced by 1.3\degr.  Using this model, the fiducial longitude then falls between the two half-core Gaussians irrespective of the leading one's relative amplitude. 

Four, in modeling B1237+25's profile and carrying out an A/R height analysis \cite{gg2003} claimed to identify a further pair of conal components interior to the outer and inner ones.  Evidence was given in Paper I contravening this claim; however, it is also true that the pulsar's full profiles show weak power at longitudes in this region of the profile's core/double-cone structure.  We believe that much of this can be traced to abnormal mode contributions to the total profile; see Fig.~\ref{fig7}.  Therefore, we have found it useful to model this putative ``more inner'' cone adjacent to the core component.

\begin{figure}
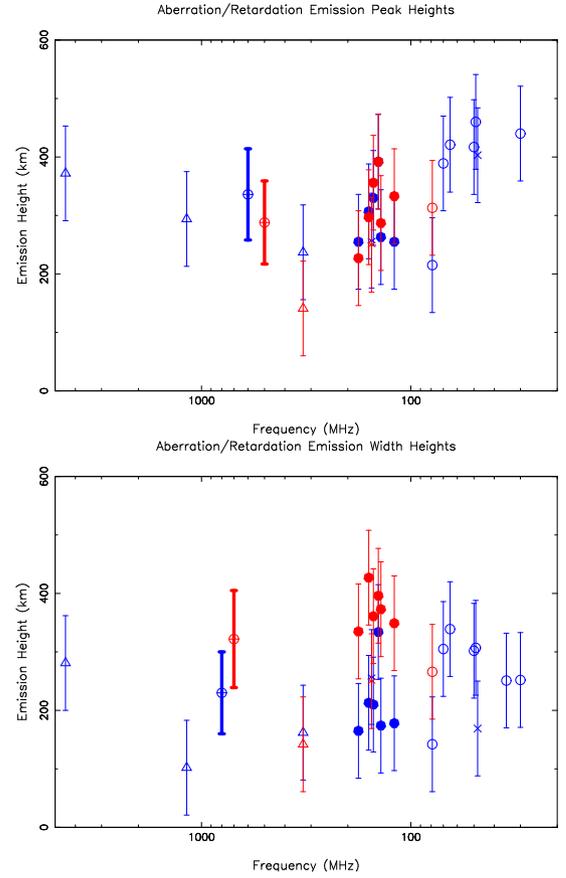

\begin{center}
\includegraphics[width=57mm,angle=-90.]{plots/B1237+25_Peak_ARhghts_rev2.ps}
\includegraphics[width=57mm,angle=-90.]{plots/B1237+25_Wdt_ARhghts_rev2.ps}
\caption{Aberration/retardation emission heights estimated from B1237+25's component peaks (top panel) and widths (bottom panel).  Those for the outer cone (2) are shown in blue and the inner (1) red.   Only values with estimated errors less than 100 km are plotted.  Symbols: AO, triangles; HBA/LBA, open/closed circles; LWA/MWA hex/times.  The corresponding average values are shown in bold in the center left of the diagram.}
\label{fig9}
\end{center}
\end{figure}

\subsection{Results of Gaussian Fitting Analyses}
The first results of our Gaussian-component modeling A/R analysis are given in Table~\ref{A/RwPeaks} and plotted in Fig.~\ref{fig9} (upper panel).
Here, $\varphi_{l/t}$ is simply $\varphi^{peak}_{l/t}$ for each outer (cone 2, blue) and inner (1, red) component pair. The observations come from a number of different instruments as shown in the second column, and are listed in increasing order of frequency.

Each total profile was fitted with the two Gaussian components used to model the core component and then pairs of Gaussian components for the outer, inner and residual pairs.  An example of this fitting is given in Fig.~\ref{fig8} using Michael Kramer's (\citeyear{kramer94}) \textsc {bfit} program. At higher frequencies where the S/N is excellent and RFI virtually absent, the fits are very good with goodness-of-fit values of only a little larger than unity---but at very high frequency the pulsar's components become conflated and the core is weak, perhaps degrading the fits.  Close inspection of Fig.~\ref{fig8} reveals the well-known problem that pulsar components tend not to have Gaussian forms at low power levels. 

Plots similar to Fig.~\ref{fig8} are given in the Appendix for each of the fitted profiles along with their corresponding residual diagrams; see Figs.~\ref{figB1} to \ref{figB18}. 

\begin{table}
\caption{Average A/R Emission Heights}
\label{A/Ravrgs}
\begin{center}
\begin{tabular}{ccc}
\toprule
    Method & Height & Fraction  \\
    Cone & (km) &  $R_{lc}=66005$ km\\
    \midrule
    \midrule	

Conal Component Peaks \\
Outer (2)  & 336$\pm$78 &  0.0051\\

Inner (1)     & 288$\pm$71 &   0.0044\\

Conal Half-power Widths   \\
Outer (2)    & 230$\pm$70 &   0.0035\\

Inner (1)     & 322$\pm$83 &  0.0049\\

\bottomrule
\end{tabular}
\end{center}
\end{table}

Fits to the outer conal component pair (cone 2), shown in the upper half of the table exhibit the well studied dramatic increase in low frequency width---and then cone radius $\rho_{lofl}^i$---but show no significant increase in emission height $h^i_{em}$.  The values fall in the 200-400 km range with an average of 336 km and rms of 78 km.  Interestingly, this later value is not different from the estimated height uncertainty based on profile measuring errors of 0.1\degr\ longitude.  The average A/R shift is about 0.5\degr\ in longitude or some 2 ms.  We find it surprising that such a small effect seems to be consistently measurable in this bright but slowish normal pulsar suggesting a remarkable solidity of the profile structure.  

Nonetheless, the A/R result is small but follows from application of a physical, not a phenomenological analysis.  The value is expectedly small for a slow pulsar and any reasonable emission height. What is remarkable is that the delays are uniformly negative and broadly consistent with each other.  Were A/R height estimates to verify what is implied from RFM, the delays would also be negative but substantially larger relative to the uncertainties.

The A/R emission heights estimated for the inner cone (1) are nearly identical with an average and rms of 288 and 71~km, respectively.  The inner cone seems to persist in the star's profiles to very low frequencies, but becomes more and more difficult to fit reliably as its relative intensity decreases.  Given that its spacing seems to be near constant or only very slightly increasing at low frequency, we have fixed its positions in many of the fits below about 70 MHz.  

Two other studies have previously used A/R to estimate the emission heights of pulsar B1237+25's cones.  \cite{gg2003} did so using GMRT observations at 318 MHz.  It was in their work that the question of a ``more inner'' cone was raised.  As discussed above we believe the appearance of such a ``cone 0'' is an artifact of the pulsar's moding, but here a useful one.  This work used the visible core center as the fiducial point, so this shifted their results by about 0.4\degr\ or 200 km in A/R height, bringing their results into rough agreement with our own.  Similarly, Paper I attempted an A/R height analysis using an Arecibo 327-MHz observation, making a similar mistake about the core structure, and obtaining very similar results again.  In their Appendix \cite{KrzeszowskiMitra2009} assembled and identified A/R shifts in a large set of mostly higher frequency B1237+25 profiles, largely anticipating our results here.

Second, a more usual manner of studying profile evolution uses the component outer half-power widths.  Such an analysis, identical in other ways to the above, is given in Table~\ref{A/Rwidths}---also plotted in Fig.~\ref{fig9} (blue points). Here, $\varphi_{l/t}=\varphi^{peak}_{l/t}\mp\varphi^{width}_{l/t}/2$ for each component pair.

Here we see also that the computed A/R emission heights are unvarying within their uncertainties of 81 km.  The average and rms of the outer conal components are 230 and 70 km respectively, whereas those for the inner cone are 322 and 83 km.  Here the outer conal $\rho^i$ values compare well with those in Table~\ref{OCLcFL} at similar frequencies.  

We cannot now know whether the use of component peaks or widths might provide more reliable and accurate results.  The average aberration/retardation height results are summarized in Table~\ref{A/Ravrgs}, and they all agree within their errors.  Each represents 0.5\% or so of the pulsar's 66,000 km velocity-of-light cylinder radius.

\section{Discussion}
\subsection{Results}
We are hardly the first to question the venerable ``radius-to-frequency mapping'' assumption.  The \cite{Hassall2012} analysis of the multifrequency profiles of a number of pulsars fails to detect the effects of A/R but finds multiple reasons to reject any simple wavelength dependence on the emission height.

Our analysis may be the first to estimate A/R effects over a broad band in a normal pulsar---in B1237+25's case a slower one---and to use the core/cone structure as an essential means of doing so.  As mentioned above, we found few good candidate pulsars on which this study can be performed because the population of core/double-cone pulsars is small and because only a few of these pulsars remain undistorted by scattering into the decameter band.  

A number of earlier A/R studies have attempted to use a core component to locate the fiducial longitude, but have done so crudely without showing that the core had this property.  Here we have been able to show that the center of B1237+25's usually asymmetric core (which we model as having two halves of different amplitudes) aligns with both the PPA inflection point and Stokes $V$ zero crossing point at 327 MHz---and very plausibly over the entire band.  That the core's overall polarization structure is similar at 1117, 150 and 70 MHz (Figs.~\ref{fig6}, \ref{fig4} and \ref{fig5}) gives credence to these properties being broad band---indeed fully into the decametric band.  

The \cite{BCW1991} theory did not envision profile structures with both conal and core emission; rather, it most readily applies to the entirely conal profile in Fig.~\ref{fig3} (right).  That the here absent core coincides with the needed PPA inflection shows that we are applying their method correctly, however the differing character of the conal and core emission could alter how our results should be interpreted.  

We note that virtually every estimated A/R emission height in Tables~\ref{A/RwPeaks} and \ref{A/Rwidths} falls in the 200-400 km range and their averages and errors for the two cones and two methods hardly distinguish them.  In particular the outer conal width average height of 230 km is comparable to the 1-GHz characteristic height in Table~\ref{OCLcFL}.  Of course, we cannot make too much of this given the large A/R height errors, however 1-GHz LoFL height values fall in the 200+ km range for hundreds of pulsars with outer cones \citep[\eg][]{Rankin2022}.

\subsection{Did Radius-to-Frequency Mapping Ever Really Make Sense?}
Given that pulsar magnetospheres are comprised of a dense electron-positron plasma and many radiative processes operate more strongly in such an environment, we can question how it is that the radio emission ever manages to escape.  Observationally, it is also ever more clear that pulsar radio emission is comprised strongly, if not predominantly, by the less easily absorbed extraordinary (X) mode.  We  can then inquire how and where the plasma frequency ever gets low enough for the radio emission to propagate through the plasma.

Let us then estimate that pulsar radio emission comes from the lowest height where a particular frequency can propagate freely.  Then using the most seemingly straightforward physical theory, the required principles would be a) the election-positron plasma frequency $f_{pe}$, b) the \citep{Goldreich69} density $\rho_{GJ}$, c) the pulsar magnetic field relation for $B$ in terms of the period and spindown, and d) that this field scales as inverse height to the third power.  Emission regions may also entail an overdensity (or multiplicity) of $N$ compared to $\rho_{GJ}$ and relativistic particle motion $\gamma$.

Putting in the constants, this all reduces to the following relationship
\begin{equation}
h = 40 f_\mathrm{GHz}^{-2/3} (N/\gamma)^{1/3} \tau_\mathrm{MY}^{-1/6}I_{45}^{1/6}  \ \ \ \mathrm{km}
\end{equation}
where, $h$ is the emission height, $f_\mathrm{GHz}$ is the frequency in GHz, $N$ is a pure number, $\tau_\mathrm{MY}$ is the spindown age in megayears, and $I_{45}$ the moment of inertia in units of $10^{45}$ g cm$^2$.\footnote{Plasma frequency: $\omega_{pe}=(8\pi n_e e^2/\gamma m_e)^{1/2}$. Goldreich-Julian density: $\rho_\mathrm{GJ}=(1/2\pi c)\ \mathbf{\Omega \cdot B}\approx B/cP$. Multiplicity $N$: $n_e=N\rho_\mathrm{GJ}/e$. Pulsar Magnetic Field: $B_0\approx(3c^3 I/8\pi^2 R^6)^{1/2}(P\dot P)^{1/2}$ at the surface; $B=B_0 (h/R)^{-3}$ elsewhere \citep{Ruderman72}---where $e$, $\gamma m_e$ and $c$ are the electron charge, relativistic mass and lightspeed; $P$ and $\dot P$ the pulsar rotation period and its first time derivative, $I$ the moment of inertia, $R$ the stellar radius, and $h$ the emission height.}

Here, this crude free propagation height $h$ does show an inverse frequency dependence, but it is much steeper than the $f^{-1/3}$ or $f^{-1/4}$ LoFL dependence  usually attributed to ``radius-to-frequency mapping"---even the $\rho$ $\propto$ $f^{-0.45}$ above for B1237+25.  That this simplified ``radio escape'' height scales with $\tau_{MY}$ so weakly also seems to agree with a broad experience in modeling the beam geometry of pulsars with different geometries.  Even the overall scaling seems broadly compatible, where if $N$ is 100-1000 and $\tau_{MY}$ perhaps 10-50 MY, then $h$ would barely fall in the range of our results, again suggesting that pulsar radio emission is largely absorbed.  

Even such a crude application of the most basic physics seems to severely undermine ``radius-to-frequency mapping''.  Did theorists have means of understanding a shallower $h$ frequency dependence?  If so we are not aware of it.  

If a frequency dependence of the emission height is not the prime-cause of the observed widening of the pulse profile of PSR~B1237+25 at low frequencies as in Fig.~\ref{fig2}, then what is? It seems likely that frequency-dependent  propagation effects in the pulsar magnetosphere provide a viable alternative interpretation. For example, refractive effects on the ordinary (O) mode are expected to introduce a frequency dependence to the profile morphology \citep{petrova2000}.

\subsection{Wherefore Core Radiation?}
Core emission components are ubiquitous in pulsar radio profiles, and they have the stunningly puzzling property that their widths reflect the angular diameter of the polar cap at the surface.  The problem is that radio emission from this region almost certainly cannot get out and be observed.  One clear reason is that the plasma density above the polar cap is so high that any radio emission would be absorbed as soon as it was emitted.  

So apparently, core emission must come from much greater heights, probably from heights not too different than those of conal emission.  That most core emission is X-mode dominated (ETXI) provides strong evidence that it has trouble escaping the magnetosphere. Indeed, if there wasn't substantial O-mode absorption it would almost certainly be predominant. 

This all said, our A/R estimates (and many others in a variety of contexts) argue for conal emission heights of about 200-300 km.  While a core component seems to mark the fiducial longitude of a pulsar's magnetic axis accurately so as to facilitate A/R analyses, we have no comparable means to estimate the core emission height directly.  However, an important clue is the intensity-dependent A/R observed in some well studied core components, B1237+25 included, also suggesting emission heights of about this amount.  Were then some maser or other mechanism amplifying the beam plasma energy to operate along the magnetic axis, then it might well excite radio emission at some 200-km or so height.  The mystery is the putative maser or other appropriate mechanism that excites the core beam with the angular width of the polar cap.

\section{Summary}
Our B1237+25 A/R analysis above appears to impugn the validity of the venerable phenomenological analysis method of average-profile morphology mechanism known as ``radius-to-frequency mapping".  Here, the result is confined to a single pulsar, however several of us have found a comparable result for B1451--68 (Bhat et al., submitted).  The population of bright core/cone pulsars is small, and both scattering and spectral turnovers compromise their use for additional future analyses.  Nonetheless, some of us are assessing other possible candidates.  

Should ``radius-to-frequency mapping" be invalidated, a broad range of other pulsar investigations would need to be reexamined and a suitable mechanism found to explain why outer conal beam radii increase with wavelength so prominently.  Such assessments and proposals are far beyond the scope of our current project.  We are then left now with Jim Cordes' half-century old ``ad hoc counterproposal .... that all frequencies are radiated from the same narrow radial range by a broad-band mechanism but that the radiated spectrum varies systematically with distance from the magnetic pole. To reproduce the average-pulse-profile observations, the frequency of the spectral peak would necessarily decrease with distance from the magnetic pole."

\section*{Acknowledgements}

Our paper has benefitted greatly from the efforts of our reviewer who provided particularly detailed and challenging comments.
Much of JMR's work was made possible by support from the US National Science Foundation (NSF) grant 18-14397. 
The Arecibo Observatory was constructed and operated by Cornell University for half a century---largely under NSF support---then by several other contractors for shorter intervals. 
Part of this work was supported by the Project: 101131928— ACME — HORIZON- INFRA-2023-SERV-01, also by the Project 6435 of the Science and Technology Center in Ukraine and by Ukrainian Program "Scientific and scientific and technical (experimental) work in the priority area "Radiophysical and optical systems for strengthening the defense capability of the state" for 2025-2026 - "Global monitoring of radio signals of natural and artificial origin of decametre-metre waves in the interests of cosmology and applied problems of defense capability" (state registration number - 0125U000879).
We especially thank our colleagues who maintain the ATNF Pulsar Catalog and the European Pulsar Network Database. This work made use of the NASA ADS astronomical data system.
This paper is partially based on data obtained using the NenuFAR radio-telescope. The development of NenuFAR has been supported by personnel and funding from: Observatoire Radioastronomique de Nan\c{c}ay, CNRS-INSU, Observatoire de Paris-PSL, Université d’Orléans, Observatoire des Sciences de l’Univers en Région Centre, Région Centre-Val de Loire, DIM-ACAV and DIM-ACAV + of Région 
Ile-de-France, Agence Nationale de la Recherche.
We acknowledge the use of the Nan\c{c}ay Data Centre computing facility (CDN – Centre de Données de Nan\c{c}ay). The CDN is hosted by the Observatoire Radioastronomique de Nan\c{c}ay (ORN) in partnership with Observatoire de Paris, Université d'Orléans, OSUC, and the CNRS. The CDN is supported by the Région Centre-Val de Loire, département du Cher. 
The Nan\c{c}ay Radio Observatory (ORN) is operated by Paris Observatory, associated with the French Centre National de la Recherche Scientifique (CNRS) and Universit\'{e}
d'Orl\'{e}ans.
This scientific work uses data obtained from \textit{Inyarrimanha Ilgari Bundara}, the CSIRO Murchison Radio-astronomy Observatory, an initiative of the Australian Government, with support from the Government of Western Australia and the Science and Industry Endowment Fund. Support for the operation of the MWA is provided by the Australian Government (NCRIS), under a contract to Curtin University administered by Astronomy Australia Limited.
This paper also makes use of data obtained by the LWA radio telescope. Construction of the LWA has been supported by the Office of Naval Research under Contract N00014-07-C-0147 and by the AFOSR. Support for operations and continuing development of the LWA1 is provided by the Air Force Research Laboratory and the National Science Foundation under grants AST-1711164 and AGS1708855. 
This paper is based (in part) on data obtained with the LOFAR telescope (LOFAR-ERIC) under the pre-cycle 0 project code Pulsars2. LOFAR \citep{vwg+13} is the Low Frequency Array designed and constructed by ASTRON. It has observing, data processing, and data storage facilities in several countries, that are owned by various parties (each with their own funding sources), and that are collectively operated by the LOFAR European Research Infrastructure Consortium (LOFAR-ERIC) under a joint scientific policy. The LOFAR-ERIC resources have benefited from the following recent major funding sources: CNRS-INSU, Observatoire de Paris and Université d'Orléans, France; BMBF, MIWF-NRW, MPG, Germany; Science Foundation Ireland (SFI), Department of Business, Enterprise and Innovation (DBEI), Ireland; NWO, The Netherlands; The Science and Technology Facilities Council, UK; Ministry of Science and Higher Education, Poland.
JP acknowledges support from the ANR (Agence Nationale de la Recherche) grant number ANR-20-CE31-0010. GW thanks the University of Manchester for Visitor Status.

\section*{Observational Data availability}

The observational data are available by contacting the authors associated with the various observatories.




%
%
\bibliography{biblio.bib}

\appendix
\onecolumn
\setcounter{figure}{0}
\renewcommand{\thefigure}{A\arabic{figure}}
\renewcommand{\thetable}{A\arabic{table}}
\setcounter{table}{0}
\renewcommand{\thefootnote}{A\arabic{footnote}}
\setcounter{footnote}{0}


\begin{table*}
\begin{flushleft}
\bf{APPENDIX A: MULTIBAND GAUSSIAN FITTING RESULTS}
\vspace{0.2in}
\end{flushleft}
\caption{B1237+25 Multiband A/R Height Analysis Using Component Peaks}
\label{A/RwPeaks}
\begin{center}
\begin{tabular}{cccccccc}
\toprule
    Freq/ & Observatory &$\varphi^i_l$ & $\varphi^i_t$ & $\varphi^i_c $ & $\rho^i$ & $\Delta t^i$ &  $h^i_{em}$    \\
    Cone & & (\degr) & (\degr) & (\degr) & (\degr) & (ms) & (km) \\
    \midrule
    \midrule	
	
4460/2 & AO & -5.3$\pm$0.1 & 4.0$\pm$0.1 &	-0.6$\pm$0.1 & 	3.7$\pm$0.5 & 	-2.48$\pm$0.54 & 	372$\pm$81 \\
1177/2 & AO & -5.4$\pm$0.1 & 4.4$\pm$0.1 &	-0.5$\pm$0.1 &	3.9$\pm$0.5 & 	-1.96$\pm$0.54 & 	294$\pm$81 \\
327/2 & AO & -6.4$\pm$0.1 &	5.6$\pm$0.1 &	-0.4$\pm$0.1 &	4.8$\pm$0.6 & 	-1.58$\pm$0.54 & 	237$\pm$81 \\
178/2 & LHB & -7.1$\pm$0.1 & 6.3$\pm$0.1 &	-0.4$\pm$0.1 &	5.3$\pm$0.7 & 	-1.70$\pm$0.54 & 	255$\pm$81 \\
159/2 & LHB & -7.4$\pm$0.1 & 6.3$\pm$0.1 &	-0.5$\pm$0.1 &	5.5$\pm$0.7 & 	-2.05$\pm$0.54 & 	307$\pm$81 \\
154/2 & MWA & -7.3$\pm$0.1 & 6.4$\pm$0.1 &	-0.4$\pm$0.1 &	5.5$\pm$0.7 & 	-1.71$\pm$0.54 & 	257$\pm$81 \\
151/2 & LHB & -7.5$\pm$0.1 & 6.4$\pm$0.1 &	-0.6$\pm$0.1 &	5.5$\pm$0.7 & 	-2.20$\pm$0.54 & 	330$\pm$81 \\
143/2 & LHB & -7.7$\pm$0.1 & 6.3$\pm$0.1 &	-0.7$\pm$0.1 &	5.6$\pm$0.7 & 	-2.61$\pm$0.54 & 	392$\pm$81 \\
139/2 & LHB & -7.5$\pm$0.1 & 6.6$\pm$0.1 &	-0.5$\pm$0.1 &	5.6$\pm$0.7 & 	-1.76$\pm$0.54 & 	263$\pm$81 \\
120/2 & LHB & -7.8$\pm$0.1 & 6.9$\pm$0.1 &	-0.4$\pm$0.1 &	5.9$\pm$0.8 & 	-1.70$\pm$0.54 & 	255$\pm$81 \\
79/2 & LWA & -8.3$\pm$0.1 &	7.6$\pm$0.1 &	-0.4$\pm$0.1 &	6.4$\pm$0.8 & 	-1.43$\pm$0.54 & 	215$\pm$81 \\
65/2 & LWA & -9.0$\pm$0.1 &	7.6$\pm$0.1 &	-0.7$\pm$0.1 &	6.6$\pm$0.0 & 	-2.81$\pm$0.54 & 	421$\pm$81 \\
50/2 & LWA & -9.4$\pm$0.1 &	7.8$\pm$0.1 &	-0.8$\pm$0.1 &	6.8$\pm$0.0 & 	-3.06$\pm$0.54 & 	460$\pm$81 \\
48/2 & LBA & -9.3$\pm$0.1 &	7.9$\pm$0.1 &	-0.7$\pm$0.1 &	6.9$\pm$0.9 & 	-2.69$\pm$0.54 & 	403$\pm$81 \\
35/2 & LWA & -10.0$\pm$0.2 & 8.2$\pm$0.2 &	-0.9$\pm$0.3 &	7.3$\pm$1.9 & 	-3.31$\pm$1.09 & 	497$\pm$163 \\
\\
70/2 & NenuFAR & -8.9$\pm$0.1 & 7.5$\pm$0.1 & -0.7$\pm$0.1 & 6.5$\pm$0.9 & -2.59$\pm$0.54 & 389$\pm$81 \\
50/2 & NenuFAR & -9.6$\pm$0.1 & 8.1$\pm$0.1 & -0.7$\pm$0.1 & 7.0$\pm$0.9 & -2.85$\pm$0.54 & 427$\pm$81 \\
30/2 & NenuFAR & -10.4$\pm$0.1 & 8.8$\pm$0.1 & -0.8$\pm$0.1 & 7.7$\pm$1.0 & -2.93$\pm$0.54 & 440$\pm$81 \\
\\													
													
4460/1 & AO & -3.0	$\pm$0.1 &	2.2	$\pm$0.1 &	-0.4$\pm$0.1 &	2.1$\pm$0.3 &	-1.61$\pm$0.54 &	241	$\pm$163 \\
1177/1 & AO & -3.2	$\pm$0.1 &	2.3	$\pm$0.1 &	-0.4$\pm$0.1 &	2.2$\pm$0.3 &	-1.69$\pm$0.54 &	253	$\pm$163 \\
327/1 & AO & -3.9	$\pm$0.1 &	3.4	$\pm$0.1 &	-0.2$\pm$0.1 &	2.9$\pm$0.4 &	-0.94$\pm$0.54 &	141	$\pm$81 \\
178/1 & LHB & -4.5	$\pm$0.1 &	3.7	$\pm$0.1 &	-0.4$\pm$0.1 &	3.2$\pm$0.4 &	-1.51$\pm$0.54 &	227	$\pm$81 \\
159/1 & LHB & -4.7	$\pm$0.1 &	3.6	$\pm$0.1 &	-0.5$\pm$0.1 &	3.3$\pm$0.4 &	-1.98$\pm$0.54 &	297	$\pm$81 \\
154/1 & MWA & -4.6	$\pm$0.1 &	3.8	$\pm$0.1 &	-0.4$\pm$0.1 &	3.4$\pm$0.5 &	-1.67$\pm$0.54 &	250	$\pm$81 \\
151/1 & LHB & -4.9	$\pm$0.1 &	3.6	$\pm$0.1 &	-0.6$\pm$0.1 &	3.4$\pm$0.5 &	-2.38$\pm$0.54 &	356	$\pm$81 \\
143/1 & LHB & -4.9	$\pm$0.1 &	3.5	$\pm$0.1 &	-0.7$\pm$0.1 &	3.4$\pm$0.5 &	-2.61$\pm$0.54 &	392	$\pm$81 \\
139/1 & LHB & -4.7	$\pm$0.1 &	3.7	$\pm$0.1 &	-0.5$\pm$0.1 &	3.4$\pm$0.5 &	-1.91$\pm$0.54 &	287	$\pm$81 \\
120/1 & LHB & -5.0	$\pm$0.1 &	3.8	$\pm$0.1 &	-0.6$\pm$0.1 &	3.5$\pm$0.5 &	-2.22$\pm$0.54 &	333	$\pm$81 \\
79/1 & LWA & -5.5	$\pm$0.1 &	4.4	$\pm$0.1 &	-0.5$\pm$0.1 &	3.9$\pm$0.5 &	-2.09$\pm$0.54 &	313	$\pm$81 \\
65/1 & LWA & -5.5	$\pm$0.1 &	4.4	$\pm$0.1 &	-0.5$\pm$0.1 &	4.0$\pm$0.5 &	-2.11$\pm$0.54 &	317	$\pm$163\\
50/1 & LWA & -5.5	$\pm$0.1 &	4.4	$\pm$0.1 &	-0.6$\pm$0.1 &	4.0$\pm$0.0 &	-2.11$\pm$0.54 &	317	$\pm$163 \\
48/1 & LBA & -5.5	$\pm$0.1 &	4.4	$\pm$0.1 &	-0.6$\pm$0.1 &	4.0$\pm$0.5 &	-2.11$\pm$0.54 &	317	$\pm$163 \\
35/1 & LWA & -5.4	$\pm$0.4 &	4.4	$\pm$0.4 &	-0.5	$\pm$0.6 &	3.9$\pm$2.1 &	 -1.84$\pm$2.17 &   275$\pm$326 \\
\\
70/1 & NenuFAR & -6.1$\pm$0.1 & 4.4$\pm$0.1 & -0.9$\pm$0.1 & 4.2$\pm$0.6 & -3.33$\pm$0.54 & 500$\pm$163 \\
50/1 & NenuFAR & -6.8$\pm$0.1 & 4.0$\pm$0.1 & -1.4$\pm$0.1 & 4.3$\pm$0.6 & -5.26$\pm$0.54 & 352$\pm$163 \\
30/1 & NenuFAR & -7.1$\pm$0.3 & 5.2$\pm$0.3 & -1.0$\pm$0.4 & 4.9$\pm$2.0 & -3.70$\pm$1.63 & 555$\pm$244 \\
\bottomrule
\end{tabular}
 
\end{center}
\begin{flushleft}
Notes: "1" and "2" correspond to the inner and outer cones, respectively.  Notes: $\varphi^i_l$ and $\varphi^i_t$ are the leading and trailing conal component positions; $\varphi^i_c$ is their center; $\rho^i$ is the conal radius, $\Delta t^i_t$ is the total A/R shift in ms; and $h^i_{em}$ the A/R emission heights and computed from eq.(1).  The measurement errors are taken to be 0.2 and 0.3\degr\ for the outer and inner conal components, respectively. \\
Observatory codes: AO, Arecibo Observatory; LHB, LOFAR High Band; MWA, Murchison Widefield Array, LWA, Long Wavelength Array; LBA, LOFAR Low Band; and NenuFAR, NenuFAR Telescope.
\end{flushleft}
\end{table*}

\begin{table*}
\caption{B1237+25 Multiband A/R Height Analysis Using Component Widths}
\label{A/Rwidths}
\begin{center}
\begin{tabular}{cccccccc}
\toprule
    Freq/ & Observatory &$\varphi^i_l$ & $\varphi^i_t$ & $\varphi^i_c $ & $\rho^i$ & $\Delta t^i$ &  $h^i_{em}$    \\
    Cone & & (\degr) & (\degr) & (\degr) & (\degr) & (ms) & (km) \\
    \midrule
    \midrule	
4460/2 & AO & -6.1$\pm$0.1 & 5.1$\pm$0.1 &	-0.5$\pm$0.1 &	4.5$\pm$0.6 &	-1.88$\pm$0.54 &	281$\pm$81 \\
1177/2 & AO & -6.1$\pm$0.1 & 5.8$\pm$0.1 &	-0.2$\pm$0.1 &	4.8$\pm$0.6 &	-0.68$\pm$0.54 &	102$\pm$81 \\
327/2 & AO & -7.2$\pm$0.1 & 6.7$\pm$0.1 &	-0.3$\pm$0.1 &	5.5$\pm$0.7 &	-1.08$\pm$0.54 &	162$\pm$81 \\
178/2 & LHB & -8.0$\pm$0.1 & 7.4$\pm$0.1 &	-0.3$\pm$0.1 &	6.2$\pm$0.8 &	-1.10$\pm$0.54 &	165$\pm$81 \\
159/2 & LHB & -8.3$\pm$0.1 & 7.6$\pm$0.1 &	-0.4$\pm$0.1 &	6.3$\pm$0.8 &	-1.42$\pm$0.54 &	213$\pm$81 \\
154/2 & MWA & -7.3$\pm$0.1 & 6.4$\pm$0.1 &	-0.4$\pm$0.1 &	5.5$\pm$0.7 &	-1.71$\pm$0.54 &	257$\pm$81 \\
151/2 & LHB & -8.4$\pm$0.1 & 7.6$\pm$0.1 &	-0.4$\pm$0.1 &	6.4$\pm$0.8 &	-1.40$\pm$0.54 &	210$\pm$81 \\
143/2 & LHB & -8.7$\pm$0.1 & 7.5$\pm$0.1 &	-0.6$\pm$0.1 &	6.4$\pm$0.8 &	-2.23$\pm$0.54 &	334$\pm$81 \\
139/2 & LHB & -8.5$\pm$0.1 & 7.9$\pm$0.1 &	-0.3$\pm$0.1 &	6.5$\pm$0.9 &	-1.16$\pm$0.54 &	174$\pm$81 \\
120/2 & LHB & -8.8$\pm$0.1 & 8.2$\pm$0.1 &	-0.3$\pm$0.1 &	6.8$\pm$0.9 &	-1.19$\pm$0.54 &	178$\pm$81 \\
79/2 & LWA & -9.4$\pm$0.1 & 8.9$\pm$0.1 &	-0.2$\pm$0.1 &	7.3$\pm$0.9 &	-0.94$\pm$0.54 &	142$\pm$81 \\
65/2 & LWA & -10.2$\pm$0.1 & 9.0$\pm$0.1 &	-0.6$\pm$0.1 &	7.6$\pm$1.0 &	-2.26$\pm$0.54 &	339$\pm$81 \\
50/2 & LWA & -10.5$\pm$0.1 & 9.5$\pm$0.1 &	-0.5$\pm$0.1 &	8.0$\pm$1.0 &	-2.04$\pm$0.54 &	307$\pm$81 \\
48/2 & LBA & -10.7$\pm$0.1 & 10.1$\pm$0.1 &	-0.3$\pm$0.1 &	8.3$\pm$1.1 & -1.13$\pm$0.54 &	169$\pm$81 \\
35/2 & LWA & -11.2$\pm$0.1 & 9.7$\pm$0.1 &	-0.8$\pm$0.1 &	8.4$\pm$1.1 &	-2.90$\pm$0.54 &	251$\pm$81 \\
\\
70/2 & NenuFAR & -9.9$\pm$0.1 & 8.9$\pm$0.1 & -0.5$\pm$0.1 & 7.5$\pm$1.0 &	-2.04$\pm$0.54 &	305$\pm$81 \\
50/2 & NenuFAR & -10.7$\pm$0.1 & 9.7$\pm$0.1 & -0.5$\pm$0.1 & 8.1$\pm$1.0 &	-1.99$\pm$0.54 &	302$\pm$81 \\
30/2 & NenuFAR & -11.2$\pm$0.1 & 10.3$\pm$0.1 & -0.4$\pm$0.1 & 8.6$\pm$1.1 &	-1.67$\pm$0.54 &	252$\pm$81 \\
									\\				
													
4460/1 & AO & -4.4$\pm$0.1 & 2.8$\pm$0.1 &	-0.8$\pm$0.1 &	2.9$\pm$0.4 &	-3.01$\pm$0.54 &	451$\pm$163 \\
1177/1 & AO & -4.7$\pm$0.1 & 3.1$\pm$0.1 &	-0.8$\pm$0.1 &	3.1$\pm$0.4 &	-2.97$\pm$0.54 &	445$\pm$163 \\
327/1 & AO & -4.8$\pm$0.1 & 4.3	$\pm$0.1 &	-0.2$\pm$0.1 &	3.6$\pm$0.5 &	-0.94$\pm$0.54 &	142$\pm$81 \\
178/1 & LHB & -5.8$\pm$0.1 & 4.6$\pm$0.1 &	-0.6$\pm$0.1 &	4.1$\pm$0.6 &	-2.23$\pm$0.54 &	335$\pm$81 \\
159/1 & LHB & -6.0$\pm$0.1 & 4.5$\pm$0.1 &	-0.7$\pm$0.1 &	4.2$\pm$0.6 &	-2.84$\pm$0.54 &	427$\pm$81 \\
154/1 & MWA & -4.6$\pm$0.1 & 3.8$\pm$0.1 &	-0.4$\pm$0.1 &	3.4$\pm$0.5 &	-1.67$\pm$0.54 &	250$\pm$81 \\
151/1 & LHB & -5.8$\pm$0.1 & 4.5$\pm$0.1 &	-0.6$\pm$0.1 &	4.1$\pm$0.6 &	-2.40$\pm$0.54 &	361$\pm$81 \\
143/1 & LHB & -5.9$\pm$0.1 & 4.5$\pm$0.1 &	-0.7$\pm$0.1 &	4.1$\pm$0.6 &	-2.64$\pm$0.54 &	396$\pm$81 \\
139/1 & LHB & -6.1$\pm$0.1 & 4.8$\pm$0.1 &	-0.6$\pm$0.1 &	4.3$\pm$0.6 &	-2.49$\pm$0.54 &	373$\pm$81 \\
120/1 & LHB & -6.5$\pm$0.1 & 5.3$\pm$0.1 &	-0.6$\pm$0.1 &	4.7$\pm$0.6 &	-2.33$\pm$0.54 &	349$\pm$81 \\
79/1 & LWA & -6.8$\pm$0.1 & 5.9	$\pm$0.1 &	-0.5$\pm$0.1 &	5.0$\pm$0.7 &	-1.77$\pm$0.54 &	266$\pm$81 \\
65/1 & LWA & -6.9$\pm$0.1 & 5.9	$\pm$0.1 &	-0.5$\pm$0.1 &	5.1$\pm$0.7 &	-1.81$\pm$0.54 &	271$\pm$163 \\
50/1 & LWA & -7.0$\pm$0.1 & 5.4	$\pm$0.1 &	-0.8$\pm$0.1 &	4.9$\pm$0.7 &	-3.10$\pm$0.54 &	465$\pm$163 \\
48/1 & LBA & -7.8$\pm$0.1 & 6.2	$\pm$0.1 &	-0.8$\pm$0.1 &	5.6$\pm$0.7 &	-3.16$\pm$0.54 &	474$\pm$163 \\
35/1 & LWA & -7.2$\pm$0.4 & 5.5	$\pm$0.4 &	-0.9$\pm$0.6 &	5.1$\pm$2.7 & -3.42$\pm$2.17 &	513$\pm$326 \\
\\
70/1 & NenuFAR & -7.1$\pm$0.1 &	5.3$\pm$0.1 &	-0.9$\pm$0.1 &	5.0$\pm$0.66 &	-3.57$\pm$0.54 &	535$\pm$163 \\
50/1 & NenuFAR & -7.8$\pm$0.1 &	4.9$\pm$0.1 &	-1.4$\pm$0.1 &	5.0$\pm$0.67 &	-5.52$\pm$0.54 &	414$\pm$163 \\
30/1 & NenuFAR & -8.1$\pm$0.4 &	6.0$\pm$0.4 &	-1.0$\pm$0.6 &	5.6$\pm$3.0 &	-4.02$\pm$2.17 &	603$\pm$326	\\												
\bottomrule
\end{tabular}

\end{center}
See Table A1 for explanatory notes. 
\end{table*}

\setcounter{figure}{0}
\renewcommand{\thefigure}{B\arabic{figure}}
\renewcommand{\thetable}{B\arabic{table}}
\setcounter{table}{0}
\renewcommand{\thefootnote}{B\arabic{footnote}}
\setcounter{footnote}{0}

\begin{figure*}
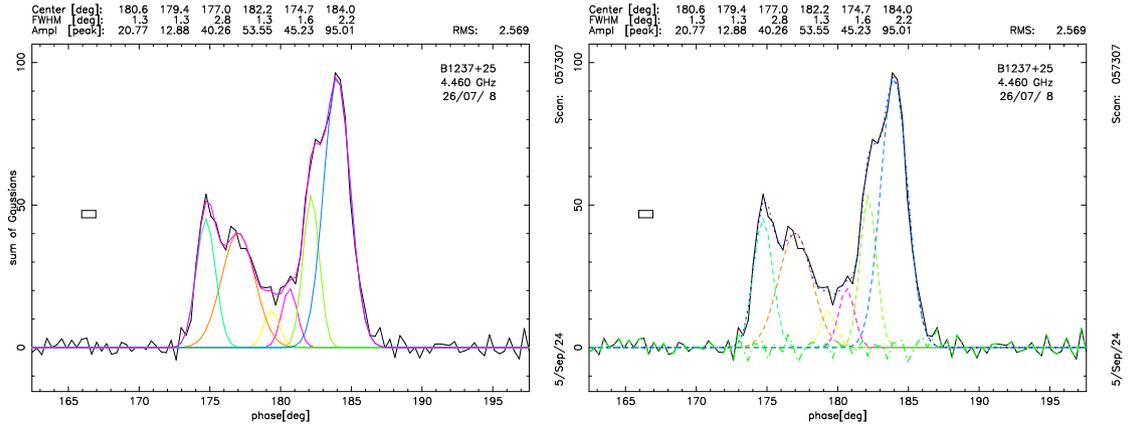

\begin{flushleft}
\bf{APPENDIX B: Gaussian Fits to Multiband Profiles}
\vspace{0.2in}
\end{flushleft}
\begin{center}
\includegraphics[width=55mm,angle=-90.]{plots/B1237+25.57307ca0_6comp.ps}
\includegraphics[width=55mm,angle=-90.]{plots/B1237+25.57307ca0_6comp_diff.ps}
\caption{Fit results for the AO 4460-MHz Observation.  The fitted composition of the profile is shown in the lefthand diagram and the residual diagram in the righthand one.}
\label{figB1}
\end{center}
\end{figure*}

\begin{figure*}
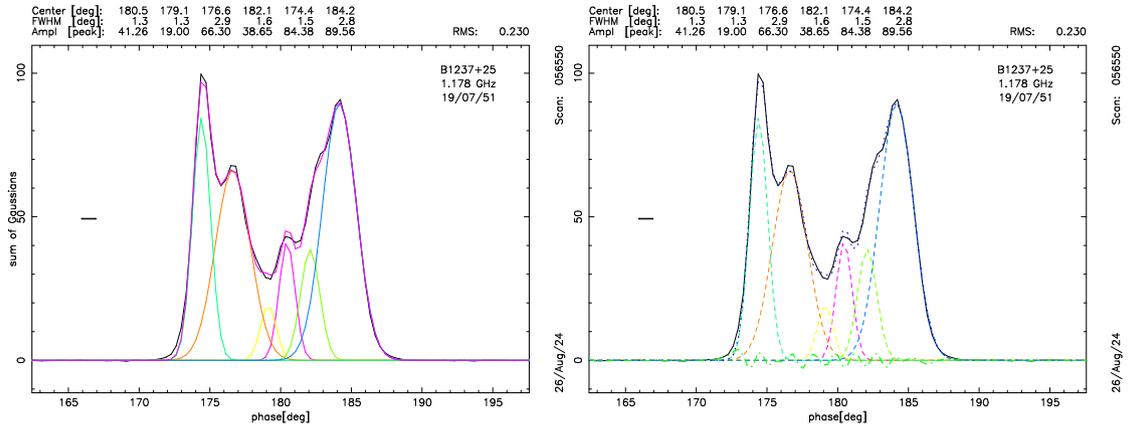

\begin{center}
\includegraphics[width=55mm,angle=-90.]{plots/pB1237+25.56550l_6comp.ps}
\includegraphics[width=55mm,angle=-90.]{plots/pB1237+25.56550l_6comp_diff.ps}
\caption{Fit results for the AO 1177-MHz Observation as in Fig.~\ref{figB1}.}
\label{figB2}
\end{center}
\end{figure*}

\begin{figure*}
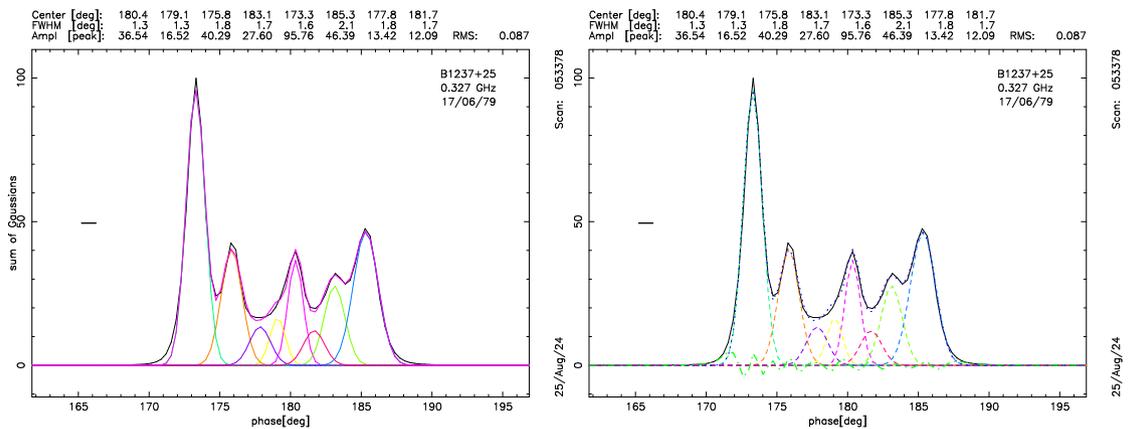

\begin{center}
\includegraphics[width=55mm,angle=-90.]{plots/pB1237+25.53378p_8comp.ps}
\includegraphics[width=55mm,angle=-90.]{plots/pB1237+25.53378p_8comp_diff.ps}
\caption{Fit results for the AO 327-MHz Observation as in Fig.~\ref{figB1}.}
\label{figB3}
\end{center}
\end{figure*}

\begin{figure*}
\begin{center}
\includegraphics[width=57mm,angle=-90.]{plots/pB1237+25_178.4MHz_8comp.ps}
\includegraphics[width=57mm,angle=-90.]{plots/pB1237+25_178.4MHz_8comp_diff.ps}
\caption{Fit results for the LHB 178-MHz Observation as in Fig.~\ref{figB1}.}
\label{figB4}
\end{center}
\end{figure*}

\begin{figure*}
\begin{center}
\includegraphics[width=57mm,angle=-90.]{plots/pB1237+25_158.7MHz_8comp.ps}
\includegraphics[width=57mm,angle=-90.]{plots/pB1237+25_158.7MHz_8comp_diff.ps}
\caption{Fit results for the LHB 159-MHz Observation as in Fig.~\ref{figB1}.}
\label{figB5}
\end{center}
\end{figure*}

\begin{figure*}
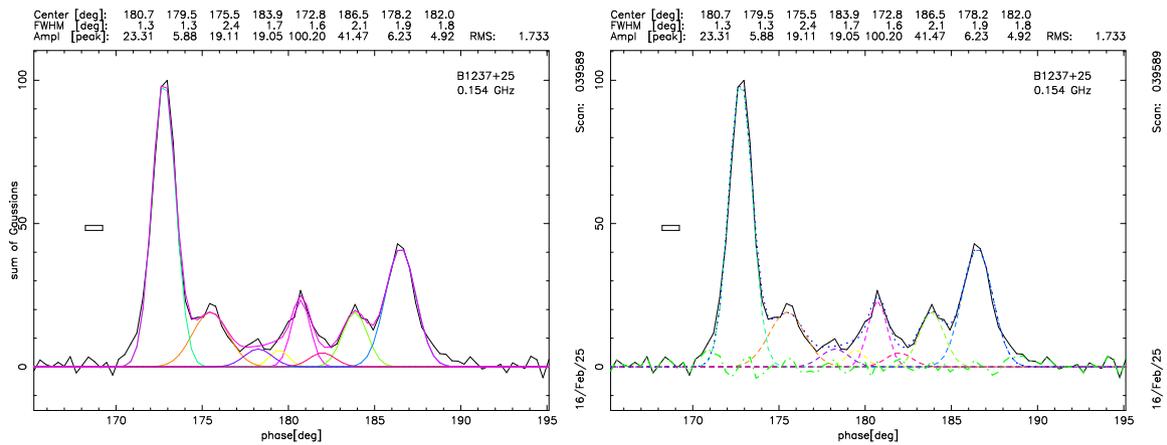

\begin{center}
\includegraphics[width=57mm,angle=-90.]{plots/pB1237+25_154.24MHz_8comp.ps}
\includegraphics[width=57mm,angle=-90.]{plots/pB1237+25_154.24MHz_8comp_diff.ps}
\caption{Fit results for the MWA 154-MHz Observation as in Fig.~\ref{figB1}.}
\label{figB6}
\end{center}
\end{figure*}

\begin{figure*}
\begin{center}
\includegraphics[width=57mm,angle=-90.]{plots/pB1237+25.78449h_8comp.ps}
\includegraphics[width=57mm,angle=-90.]{plots/pB1237+25.78449h_8comp_diff.ps}
\caption{Fit results for the LHB 151-MHz Observation as in Fig.~\ref{figB1}.}
\label{figB7}
\end{center}
\end{figure*}

\begin{figure*}
\begin{center}
\includegraphics[width=57mm,angle=-90.]{plots/pB1237+25_142.95MHz_8comp.ps}
\includegraphics[width=57mm,angle=-90.]{plots/pB1237+25_142.95MHz_8comp_diff.ps}
\caption{Fit results for the LHB 143-MHz Observation as in Fig.~\ref{figB1}.}
\label{figB8}
\end{center}
\end{figure*}

\begin{figure*}
\begin{center}
\includegraphics[width=57mm,angle=-90.]{plots/pB1237+25_139.3MHz_8comp.ps}
\includegraphics[width=57mm,angle=-90.]{plots/pB1237+25_139.3MHz_8comp_diff.ps}
\caption{Fit results for the LHB 139-MHz Observation as in Fig.~\ref{figB1}.}
\label{figB9}
\end{center}
\end{figure*}

\begin{figure*}
\begin{center}
\includegraphics[width=57mm,angle=-90.]{plots/pB1237+25_119.6MHz_8comp.ps}
\includegraphics[width=57mm,angle=-90.]{plots/pB1237+25_119.6MHz_8comp_diff.ps}
\caption{Fit results for the LHB 120-MHz Observation as in Fig.~\ref{figB1}.}
\label{figB10}
\end{center}
\end{figure*}

\begin{figure*}
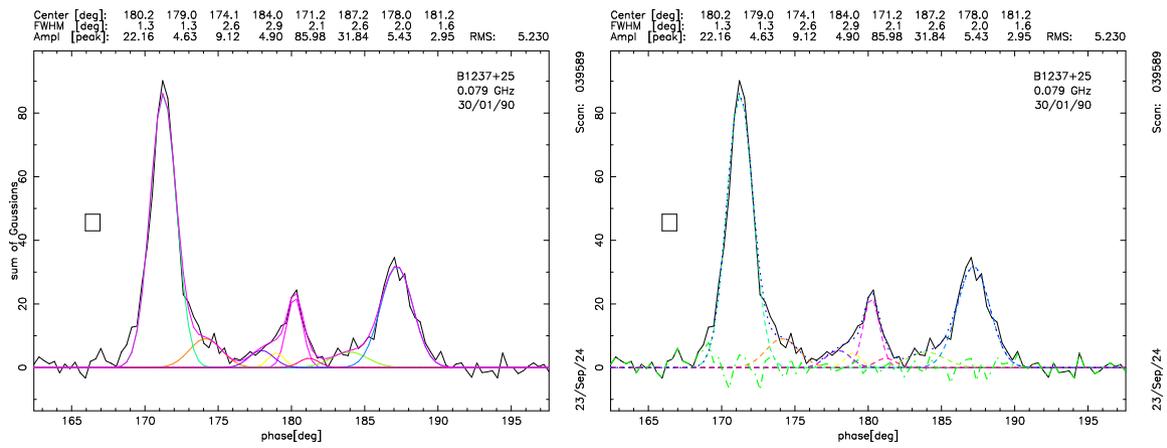

\begin{center}
\includegraphics[width=57mm,angle=-90.]{plots/B1237+25_79.2MHz_8comp.ps}
\includegraphics[width=57mm,angle=-90.]{plots/B1237+25_79.2MHz_8comp_diff.ps}
\caption{Fit results for the LWA 79-MHz Observation as in Fig.~\ref{figB1}.}
\label{figB11}
\end{center}
\end{figure*}

\begin{figure*}
\begin{center}
\includegraphics[width=57mm,angle=-90.]{plots/B1237+25_64.5MHz_B_8comp.ps}
\includegraphics[width=57mm,angle=-90.]{plots/B1237+25_64.5MHz_B_8comp_diff.ps}
\caption{Fit results for the LWA 65-MHz Observation as in Fig.~\ref{figB1}.}
\label{figB12}
\end{center}
\end{figure*}

\begin{figure*}
\begin{center}
\includegraphics[width=57mm,angle=-90.]{plots/B1237+25_49.8MHz_B_8comp.ps}
\includegraphics[width=57mm,angle=-90.]{plots/B1237+25_49.8MHz_B_8comp_diff.ps}
\caption{Fit results for the LWA 50-MHz Observation as in Fig.~\ref{figB1}.}
\label{figB13}
\end{center}
\end{figure*}

\begin{figure*}
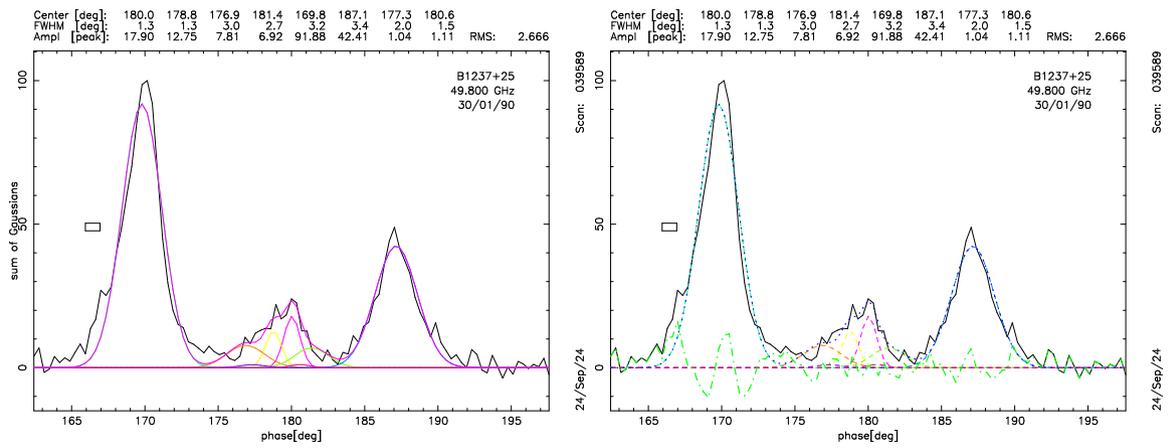

\begin{center}
\includegraphics[width=57mm,angle=-90.]{plots/B1237+25_48.4MHz_LBA_8comps.ps}
\includegraphics[width=57mm,angle=-90.]{plots/B1237+25_48.4MHz_LBA_8comps_diff.ps}
\caption{Fit results for the LWA 48-MHz Observation as in Fig.~\ref{figB1}.}
\label{figB14}
\end{center}
\end{figure*}

\begin{figure*}
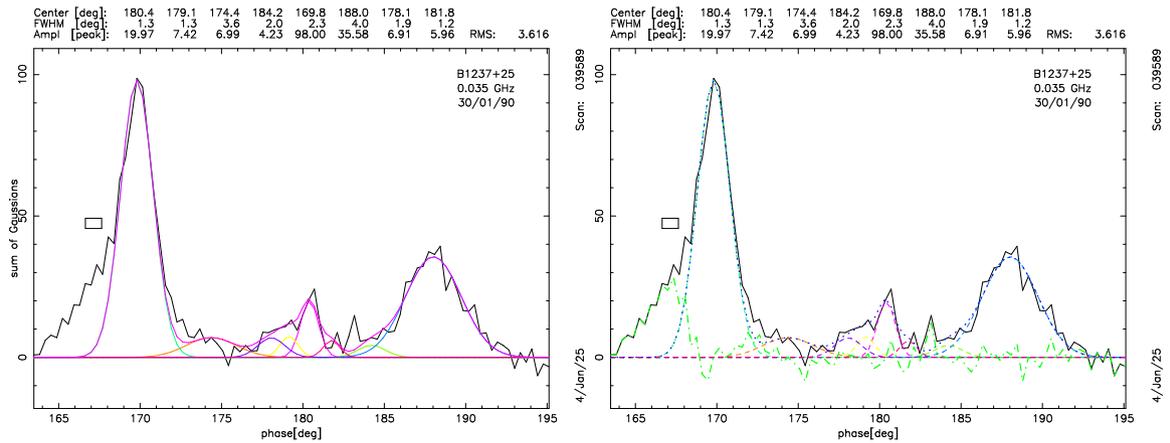

\begin{center}
\includegraphics[width=57mm,angle=-90.]{plots/B1237+25_35.1MHz_8comp_B.ps}
\includegraphics[width=57mm,angle=-90.]{plots/B1237+25_35.1MHz_8comp_diff_B.ps}
\caption{Fit results for the LWA 35-MHz Observation as in Fig.~\ref{figB1}.  The non-Gaussian shape of the leading component makes this profile impossible to fit and measure 
accurately.  It was necessary to adjust the values of components 1, 2 and 4 to obtain compatible values.}
\label{figB15}
\end{center}
\end{figure*}

\begin{figure*}
\begin{center}
\includegraphics[width=57mm,angle=-90.]{plots/B1237+25_70MHz_8comp.ps}
\includegraphics[width=57mm,angle=-90.]{plots/B1237+25_70MHz_8comp_diff.ps}
\caption{Fit results for the NenuFAR 70-MHz Observation as in Fig.~\ref{figB1}.}
\label{figB16}
\end{center}
\end{figure*}

\begin{figure*}
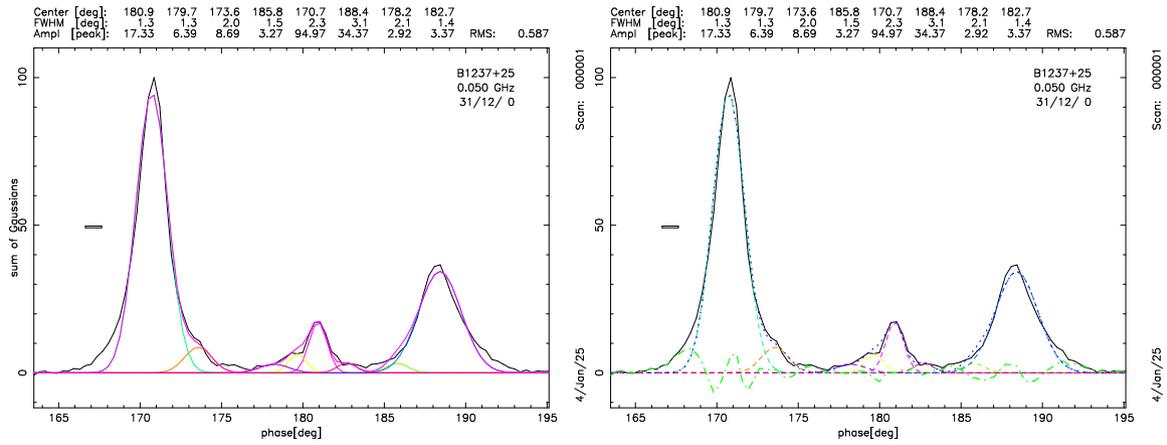

\begin{center}
\includegraphics[width=57mm,angle=-90.]{plots/B1237+25_50MHz_8comp_B.ps}
\includegraphics[width=57mm,angle=-90.]{plots/B1237+25_50MHz_8comp_diff_B.ps}
\caption{Fit results for the NenuFAR 50-MHz Observation as in Fig.~\ref{figB1}.  Positions of cones 1 and 0 trailing positions fixed.}
\label{figB17}
\end{center}
\end{figure*}

\begin{figure*}
\begin{center}
\includegraphics[width=57mm,angle=-90.]{plots/B1237+25_30MHz_8compB.ps}
\includegraphics[width=57mm,angle=-90.]{plots/B1237+25_30MHz_8comp_diffB.ps}
\caption{Fit results for the NenuFAR 30-MHz Observation as in Fig.~\ref{figB1}.}
\label{figB18}
\end{center}
\end{figure*}



\end{document}